\documentclass[fleqn,usenatbib]{mnras}

\usepackage{newtxtext,newtxmath}

\usepackage[T1]{fontenc}
\usepackage{ae,aecompl}

\usepackage{graphicx}	
\usepackage{amsmath}	
\usepackage{amssymb}	
\usepackage{times}
\PassOptionsToPackage{hyphens}{url}
\usepackage{color}
\definecolor{Brown}{rgb}{0.647,0.165,0.165}
\definecolor{NavyBlue}{rgb}{0.0,0,0.5}
\definecolor{Burgundy}{rgb}{0.5,0.0,0.125}
\hypersetup{
    breaklinks=true,
    colorlinks=true,
    citecolor={Burgundy},
    linkcolor={NavyBlue},
    urlcolor={Brown}
}

\def\apj{Astrophys.\ J.}

\def\mnras{Mon.\ Not.\ R.\ Astron.\ Soc.}
\def\aap{Astron.\ Astrophys.}
\def\apjl{Astrophys.\ J.\ Lett.}

\def\physrep{Phys.\ Rep.}
\def\pre{Phys.\ Rev.\ E}
\def\prl{Phys.\ Rev.\ Lett.}
\def\prd{Phys.\ Rev.\ D}
\def\apjs{ApJS}

\def\nat{Nature}

\def\ssr{SSRv}

\def\araa{Ann.\ Rev.\ Astron.\ Astrophys.}

\newcommand{\lb}{\ell_b} 

\newcommand{\cs}{c_\mathrm{s}}           
\newcommand{\Mach}{\mathcal{M}}      
\newcommand{\MachA}{\mathcal{M}_{\text A}}      
\newcommand{\Pm}{\text{Pm}}
\renewcommand{\Re}{\text{Re}} 
\newcommand{\Rm}{\text{Rm}} 
\newcommand{\Rmc}{\text{Rm}^{\text{(crit)}}}    

\renewcommand{\vec}[1]{\mathbf{#1}}	
\newcommand{\dd}{\mathrm{d}}        

\newcommand{\cm}{\,{\rm cm}}    
\newcommand{\km}{\,{\rm km}}    
\newcommand{\pc}{\,{\rm pc}}     
\newcommand{\kpc}{\,{\rm kpc}}  
\newcommand{\Mpc}{\,{\rm Mpc}}  

\newcommand{\g}{\,{\rm g}}      

\newcommand{\s}{\,{\rm s}}      
\newcommand{\yr}{\,{\rm yr}}    

\newcommand{\kms}{\km\s^{-1}}    

\newcommand{\G}{\,{\rm G}}      
\newcommand{\muG}{\mu{\rm G}} 

\newcommand{\K}{\,{\rm K}}      

\newcommand{\Brms}{\,B_{\rm rms}}
\newcommand{\brms}{\,b_{\rm rms}}

\newcommand{\urms}{\,u_{\rm rms}}

\newcommand\Eq[1]{Eq.\,\ref{#1}}
\newcommand\Fig[1]{Fig.~\ref{#1}}
\newcommand\Sec[1]{Sec.~\ref{#1}}
\newcommand\Tab[1]{Table~\ref{#1}}

\graphicspath{{./}{figs/}}

\title[Seed magnetic fields and small-scale dynamos]{Seed magnetic fields in turbulent small-scale dynamos}

\author[Seta and Federrath]{
Amit Seta 
\thanks{E-mail: amit.seta@anu.edu.au}
and Christoph Federrath
\\
Research School of Astronomy and Astrophysics, 
Australian National University, Canberra, ACT 2611, Australia\\
}

\date{Accepted XXX. Received YYY; in original form ZZZ}

\pubyear{2020}

\begin{document}
\label{firstpage}
\pagerange{\pageref{firstpage}--\pageref{lastpage}}
\maketitle

\begin{abstract}
Magnetic fields in galaxies and galaxy clusters are amplified from a very weak seed value to the observed
$\mu{\rm G}$ strengths by the turbulent dynamo. The seed magnetic field can be of primordial or astrophysical origin.
The strength and structure of the seed field, on the galaxy or galaxy cluster scale, can be very different,
depending on the seed-field generation mechanism. The seed field first encounters the small-scale dynamo, thus we investigate the effects of the strength and structure of the seed field on the small-scale dynamo action. 
Using numerical simulations of driven turbulence and considering three different seed-field configurations: 1) uniform field, 2) random field with a power-law spectrum, and 3) random field with a parabolic
spectrum, we show that the strength and statistical properties of the dynamo-generated magnetic fields are independent of the details of the seed field. We demonstrate that, even when the small-scale dynamo is not active, small-scale magnetic fields can be generated 
and amplified linearly due to the tangling of the large-scale field. In the presence of the small-scale dynamo action,
we find that any memory of the seed field for the non-linear small-scale dynamo generated magnetic fields is lost and thus, it is not
possible to trace back seed-field information from the evolved magnetic fields in a turbulent medium.
\end{abstract}




\section{Introduction} \label{intro}
Magnetic fields are observed in astrophysical systems over a large range of scales, 
starting from solar-system planets to the largest gravitationally bound structures, clusters of galaxies. Magnetic fields, in most of these systems, are believed to be amplified and then maintained by a turbulent dynamo mechanism, which is the key mechanism to convert turbulent kinetic energy into magnetic energy \citep{Widrow2002,BrandenburgS2005,KulsrudZ08,Federrath2016,Rincon2019}. The exponential growth of the magnetic field caused by dynamo action is determined by the induction equation. However, the induction equation does not have a source term for magnetic fields. Instead, it can only amplify a pre-existing magnetic field. Thus, one requires a seed field, even if it is very weak, to explain the presence of magnetic fields in astrophysical objects. The turbulent dynamo then amplifies the weak seed field, to produce the $\muG$ field strength (magnetic energy in near equipartition with the turbulent kinetic energy) observed in galaxies \citep{Beck2016} and galaxy clusters today \citep{CarilliT2002,GovoniF2004}. The near equipartition (depending on the parameters of the study and the nature of the turbulent driving) magnetic fields are also shown recently via small-scale turbulent dynamo experiments with laser-induced plasma turbulence \citep{TzeferacosEA2018}.

The origin of the seed magnetic fields may be primordial or astrophysical. Primordial
magnetic fields can be generated in the early Universe during inflation or various phase transitions \citep{WidrowEA2012,DurrerN2013,Subramanian2016}.
However, whether such a field survives to act as a seed for the dynamo in galaxies and galaxy clusters is still not clear. Even if such a field 
survives, it can only serve as a seed field, because primordial fields cannot explain the observed field strengths in the present-day galaxies and galaxy clusters (primordial fields decay due to turbulent diffusion and also by winds in case of galaxies \footnote{In spiral galaxies, the magnetic field is also lost via a flux expulsion mechanism \citep{Weiss1966,Moffatt1978,GilbertMT2016} however, this turns out to be less efficient compared to the decay of magnetic fields via turbulent diffusion and galactic winds \citep[Sec. 1.5 in][]{Seta2019}.}, and they might also evolve and generate turbulence; \cite{BrandenburgKT2015}).
Astrophysical seed fields can be generated due to the Biermann battery mechanism
\citep[generation of weak magnetic fields due to charge separation between electrons and ions;][]{Biermann1950},
due to ejecta from astrophysical bodies, and through plasma instabilities. 
The Biermann battery mechanism can occur during cosmic reionization \citep{SubramanianNC1994,GnedinFZ2000,LangerM2018},
during the formation of the large-scale structure \citep{KulsrudEA1997a}, and in proto-galaxies \citep{KulsrudEA1997b,DaviesW2000}. 
Fields expelled from astrophysical objects, such as from the first stars and from active galactic nuclei (AGN), can also act as astrophysical seed fields \citep{Rees2005}.
The Weibel instability, which is the generation of magnetic fields due to counter-streaming plasma, can also give rise to seed fields \citep{SchlickeiserS2003}.

Given this variety of mechanisms, it is clear that the seed fields generated by different processes have very different strengths and coherence lengths \citep{Subramanian2019}.
For example, on the scale of a galaxy ($\sim 10 \kpc$) or galaxy cluster ($\sim \Mpc$), the seed field generated during inflation or structure formation
can be fairly uniform or can have smaller correlation lengths of the order of $1 \kpc$ \citep[this depends on the inflation model, see][and references therein]{SharmaEA2017, SharmaSS2018}, whereas seed fields generated by the Weibel instability have an extremely small coherence length ($\sim$ ion skin-depth $\sim 10^{-8} \pc$). It is not clear whether these different seed fields might have an effect on the final amplified fields, in particular with respect to their amplitude and spatial distribution. In this paper, we therefore study the role of the strength and structure of the seed magnetic fields on the turbulent small-scale dynamo.

\subsection{Turbulence and dynamos}

In galaxies, turbulence can be driven on a wide range of scales, and by a variety of physical mechanisms \citep{MacLowK2004,Elmegreen2009,KrumholzB2016,FederrathEA2017,KrumholzEA2018}.
On large scales, interstellar turbulence is likely driven by supernova explosions with a 
driving scale $l_0\sim 100 \pc$, roughly equal to the size of a supernova remnant
\citep[also seen in observations;][]{HaverkornEA2008}. In case of galaxy clusters, the turbulence can be driven by merger shocks, galaxy motions, and AGN outflows and the cluster magnetic field can also be explained by the small-scale dynamo action \citep{RuzmaikinSS1989,SchekochihinC2006}.

In spiral galaxies, the observed magnetic field, based on the driving scale of turbulence $l_0$, is usually divided into small-scale (correlation length $<l_0$) and large-scale 
(correlation length $>l_0$; usually a few $\kpc$) magnetic fields. Consequently, dynamos are also divided into small-scale and large-scale dynamos, generating
small-scale and large-scale magnetic fields, respectively. The amplification due to the small-scale dynamo is physically explained by stretching 
\citep{VainshteinZ1972,SchekochihinEA2004,SetaBS2015,SetaEA2020} or compressing \citep{FederrathEA2011,FederrathEA2014,Federrath2016} 
of magnetic field lines. The small-scale dynamo quickly amplifies the weak seed field within a few turbulent turnover times ($\sim10^{7} \yr$ in spiral galaxies) to strengths comparable to 
the turbulent kinetic energy. The dynamo then saturates due to back-reaction by the Lorentz force. 
When the saturation has happened, the small-scale dynamo maintains the current energy level of the magnetic field.

After the small-scale dynamo has saturated, the large-scale (or mean-field) dynamo can order 
the small-scale dynamo amplified magnetic field into an even stronger large-scale magnetic field \citep{RuzmaikinSS1988,BeckEA1996}. 
This large-scale dynamo requires large-scale differential rotation and/or density stratification \citep{KrauseR1980,RuzmaikinSS1988,BeckEA1996,ShukurovS2008}, 
both of which are general features of accretion discs, such as proto-stellar and proto-planetary discs, as well as galactic discs.

Since the seed field first encounters the small-scale dynamo, we here focus on determining the role of the strength and structure of the seed field for the turbulent small-scale dynamo.

\subsection{Previous work on the small-scale turbulent dynamo}

The small-scale dynamo has been studied extensively, both in theoretical works 
\citep{Batchelor1950,Kazantsev1968,ZeldovichRS1990,KulsrudA1992,Subramanian1999,Subramanian2003,BoldyrevC2004,SchekochihinK2001,
SchoberEA2012a,SchoberEA2012b,BhatS2014,SchoberEA2015,AfonsoMV2019}
and via idealised numerical simulations \citep{MeneguzziFP1981,Cattaneo1999,HaugenBD2004,SchekochihinEA2004,ChoR2009,CattaneoT2009,
BushbyPW2010,FederrathEA2011,FavierB2012,SurEA2012,Beresnyak2012,BhatS2013,Cho2014,FederrathEA2014,BushbyF2014,SurBS2018,SetaEA2020}. Recent cosmological simulations also show signatures of the small-scale dynamo action in galaxies \citep{PakmorEA2017,RiederT2016,RiederT2017,RiederT2017b} 
and galaxy clusters \citep{VazzaEA2018,MarinacciEA2018,DominguezEA2019}.

Most studies tend to explore the strength, spectra, and structural properties of the dynamo-generated fields and their coupling with the turbulent velocity fields.
The main conclusions from previous works are that, if the magnetic Reynolds number ($\Rm$) is greater than its critical value (which depends on the properties of the velocity flow and magnetic resistivity)
\begin{itemize}
\item the magnetic field first grows exponentially (kinematic stage) from the seed field and then saturates (saturated stage),
\item the magnetic field power spectra seem to follow a power law with an exponent $3/2$ in the kinematic stage \citep{Kazantsev1968},
\item the magnetic field is more volume filling in the saturated stage as compared to the kinematic stage.
\end{itemize}
All previous simulations consider either a uniform or tangled seed magnetic field. Here we present a systematic study of the effects of different strengths and structures of the seed magnetic field.

In \Sec{num}, we describe our numerical model, and in \Sec{res} we determine the effects of the strength and structure of the seed
magnetic field on the turbulent small-scale dynamo. We discuss our results in \Sec{dis}, and conclude in \Sec{con}.

\section{Simulation methods} \label{num}

\subsection{Basic equations}

We solve the equations of compressible magnetohydrodynamics (MHD) in three dimensions for an isothermal gas using a modified version of the 
FLASH code \citep{FryxellEA2000,DubeyEA2008}, version 4, on a uniform, triply periodic grid with a constant viscosity and resistivity,
\begin{align}
&\frac{\partial \rho}{\partial t} + \nabla \cdot (\rho \vec{u}) = 0,  \label{ce} \\
&\frac{\partial (\rho \vec{u})}{\partial t} + \nabla \cdot \left(\rho \vec{u} \otimes \vec{u} - \frac{1}{4 \pi} \vec{B} \otimes \vec{B}\right) + 
\nabla \left(\cs^2 \rho + \frac{\vec{B}^2}{8 \pi}\right) =  \nonumber \\ 
& \hspace{0.675\columnwidth} \nabla \cdot (2 \nu \rho \vec{S}) + \rho \vec{F}, \label{ns} \\
&\frac{\partial \vec{B}}{\partial t} = \nabla \times (\vec{u} \times \vec{B}) + \eta \nabla^2 \vec{B},  \label{ie} \\
&\nabla \cdot \vec{B} = 0 \label{div},
\end{align}
where $\rho$ is the gas density, $\vec{u}$ is the velocity, $\vec{B}$ is the magnetic field, $\cs$ is the constant speed of sound, $\nu$
is the viscosity, $\eta$ is the resistivity, $S_{ij} = (1/2) \left(v_{i,j} + v_{j,i} - (2/3) \delta_{ij} \nabla \cdot \vec{v} \right)$ is the rate of strain tensor and
$\vec{F}$ is the turbulent acceleration field. Throughout the paper, the total field is referred to as $\vec{B}$, 
the mean field as $\vec{B_0}$, and the small-scale random field as $\vec{b}$.

\subsection{MHD solver, turbulent driving, and Reynolds numbers}

We use the HLL3R Riemann solver \citep{BouchutKW2007,BouchutKW2010,WaaganFK2011} to solve the above equations.
We drive turbulence in the numerical domain of size $L$ with $512^3$ grid points using $\vec{F}$ (in \Eq{ns}) modelled as an Ornstein-Uhlebeck process in Fourier space \citep{EswaranPope1988,FederrathEA2010}. The flow is driven at wavenumbers $1 \le kL/2\pi \le 3$ 
(the wavenumbers in each direction $k_x, k_y,$ and $k_z$ are chosen to be integers to ensure periodicity but the resulting $k =(k_x^2 + k_y^2 + k_z^2)^{1/2}$
need not be an integer) with a parabolic function for the power that peaks at $kL/2\pi = 2$ and reaches zero at $kL/2\pi = 1$ and $kL/2\pi = 3$. 
Thus, the driving scale of the turbulence is approximately equal to $L/2$.
For turbulence driven by supernova explosions in the ISM of spiral galaxies, the driving length scale
is roughly around $100 \pc$. We select the auto-correlation timescale for $\vec{F}$ as the eddy turnover time $t_0$ of the flow 
with respect to the driving scale, i.e., $t_0 = (L/2)/\urms$, where $\urms$ is the root mean square of the velocity field. 
In the ISM of spiral galaxies,  $\urms\simeq10\kms$ \citep{MacLowK2004,Shukurov2004} and thus $t_0 \approx 50 \pc/10 \kms \simeq 0.5 \times 10^{7} \yr$.
We drive only solenoidal modes and chose the strength of the forcing function to reach a (statistically steady) turbulent Mach numbers $\Mach=\urms/\cs=0.1$ and $10$.
At a given Mach number, solenoidal modes maximize the efficiency of the small-scale dynamo and
the growth rate decreases on increasing the Mach number \citep{FederrathEA2011,FederrathEA2014,AfonsoMV2019}. 
We use the same forcing function for all our simulations.

We define the Reynolds number ($\Re$) and the magnetic Reynolds number ($\Rm$) in terms of the rms velocity $\urms$ and the driving scale ($L/2$) of 
turbulence, as
\begin{align}
\Re = \frac{\urms L}{2 \nu} \quad \text{and} \quad \Rm = \frac{\urms L}{2\eta}. \label{rm}
\end{align}
These values then determine the magnetic Prandtl number $\Pm=\Rm/\Re$. For our dynamo simulations, we set $\Pm\ge1$ to maximize the efficiency of the small-scale  
dynamo action \citep{BoldyrevC2004,IskakovEA2007,FederrathEA2014,BrandenburgEA2018}. 
The critical magnetic Reynolds $\Rmc$ above which the magnetic fields grows exponentially is approximately $165 \, \Pm^{-1/2}$ at $\Mach=0.1$ \citep{HaugenBD2004}.
We choose $\Re=2000 (\nu = 2.5 \times 10^{-5})$ and $\Rm = 2000 \, (\eta = 2.5 \times 10^{-5}), 3000 \, (\eta = 1.67 \times 10^{-5}), $ and $100 \, (\eta = 5 \times 10^{-4}; \Rm < \Rmc)$ for our runs.

\begin{figure*}
\includegraphics[width=1\columnwidth]{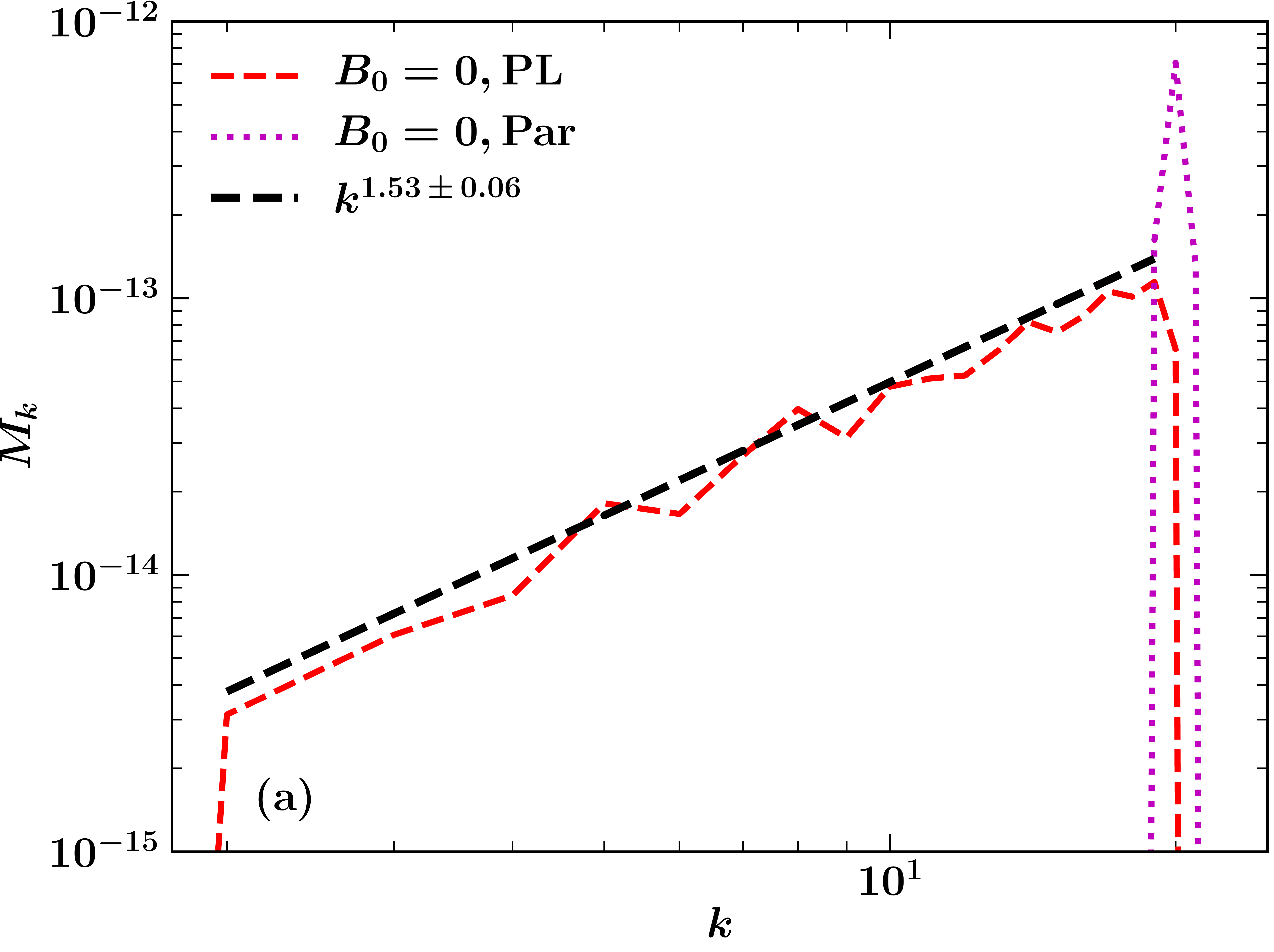} \hspace{0.5cm}
\includegraphics[width=1\columnwidth]{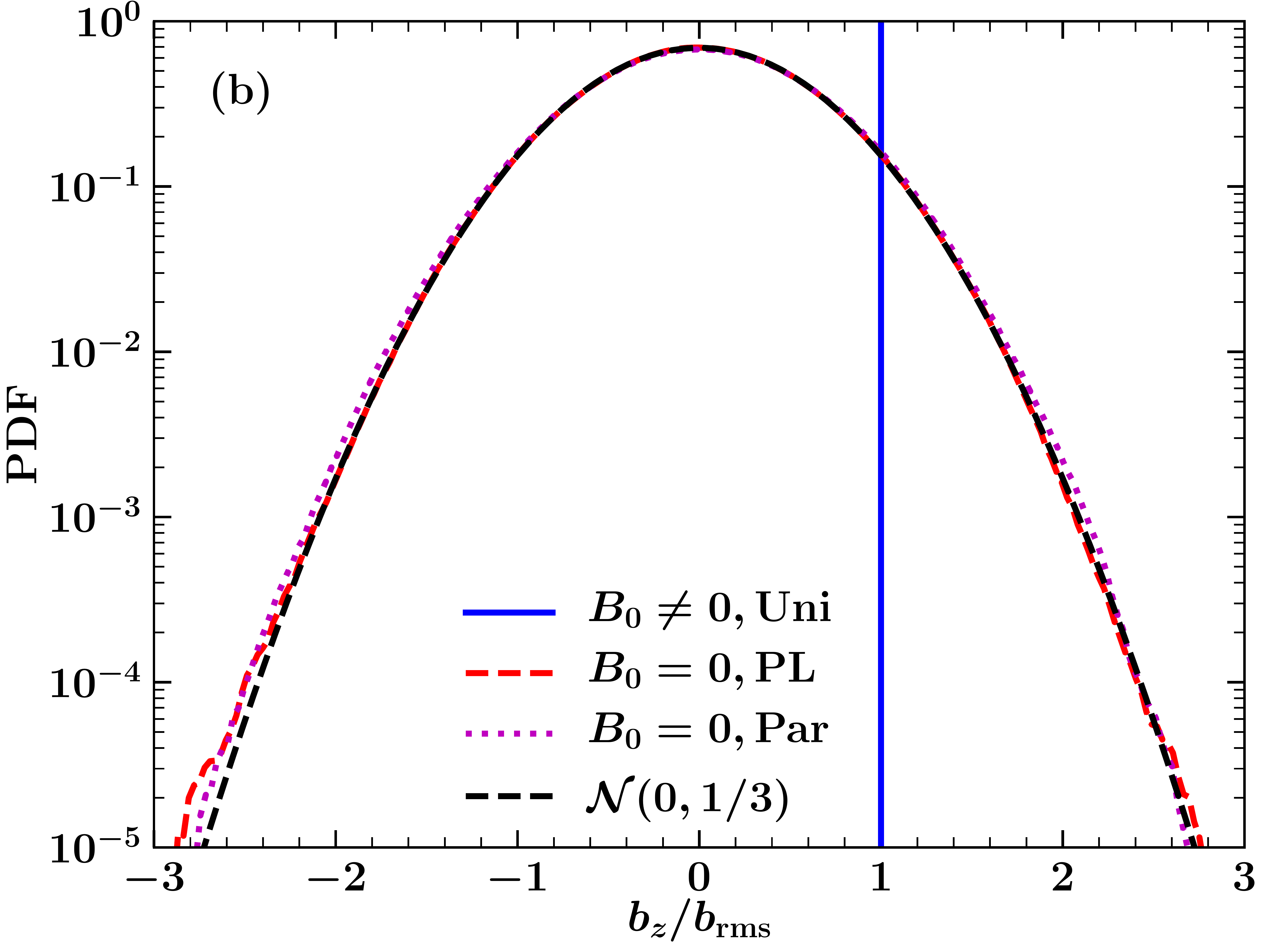}
\caption{(a) Magnetic power spectrum for random seed fields (with the mean field $B_0=0$): the seed field with a power-law ($B_0=0$, PL) spectrum 
having a slope of $3/2$ (dashed, black) in the range of wave numbers $2 \le kL/2 \pi \le 20$ (dashed, red) and the seed field with a localised (in $k$ space) 
parabolic ($B_0=0$, Par) spectrum in the range $19 \le kL/2 \pi \le 21$ with a peak at $k L/2\pi = 20$ (dotted, magenta). (b) The probability distribution function (PDF) of 
the normalised $z$ component $b_z/\brms$ of the seed magnetic field for all three cases: uniform magnetic field $B_0\ne0$, Uni (solid, blue), $B_0=0$, PL (dashed, red), 
and $B_0=0$, Par (dotted, magenta). The random fields follow a Gaussian distribution $\mathcal{N}$ with mean $0$ and standard deviation $1$ (dashed, black).}
\label{inicond}
\end{figure*}

\subsection{Seed fields, initial conditions, and simulation parameters}

Initially, we prescribe a constant density $\rho \, (t=0)$ and $\vec{u}=0$ at each grid point in the numerical domain. The density does not change much throughout
the simulation for the low-Mach number simulation ($\Mach=0.1$). In contrast, the supersonic run ($\Mach=10$) develops strong shocks and steep density gradients \citep{Federrath2013}.

\begin{table*}
	\caption{Summary of the main simulation parameters. Note that all simulations are done in a computational domain of size $L^3$ and with $512^3$ mesh points.
        In all cases, the flow is driven on large scales ($1 \le kL/2 \pi \le 3$)  and the fluid Reynolds number is set to be $\Re = \urms L / (2 \nu) = 2000$,
        where $\urms$ is the rms velocity and $\nu$ is the viscosity. 
        The columns in the table are as follows: 
        1. simulation name, 
        2. type of seed field,
        3. power spectrum of the seed field, $M_k (t=0)$, where $N$ is a constant chosen to achieve the required seed field strength (or initial Alfv\'en Mach number) and for the case of the parabolic seed field, the power peaks at the critical wave number, $k_{\rm c} = 20 \, (2 \pi / L)$,
        4. Sonic Mach number, $\Mach = \urms/\cs$, where $\cs$ is the sound speed,
        5. Alfv\'en Mach number, $\MachA (t=0) = (4\pi\rho \, (t=0))^{1/2} \urms / \Brms (t=0)$, where $\rho \, (t=0)$ is the initial constant density and $\Brms (t=0)$ is the initial rms
        total magnetic field strength,
        6. initial plasma beta, $\beta (t=0) = (p_{\rm thermal}/p_{\rm magnetic})_{t=0} = \cs^2 \rho \, (t=0) / (\Brms^2 (t=0) / 8 \pi)$,
        7. magnetic Reynolds number, $\Rm = \urms L / (2 \eta)$, where $\eta$ is the resistivity, 
        8. the computed exponential growth rate (except for run no. 13 with $\Rm < \Rmc$ for which the growth is linear), $\Gamma$, in units of $1/t_0$ ,         
        9. the final statistically saturated magnetic field energy with respect to the kinetic energy, $E_{\rm m}/E_{\rm k}$, and
        10. the final statistically saturated amplification factor, $E_{\rm m}/E_{\rm m0}$.
        The errors reported in the calculated quantities (8., 9., and 10.) represent the maximum of the computed fitting and systematic errors for each case.}
	\label{param}
	\begin{tabular}{lccccccccc} 
		\hline 
		\hline
		Simulation name & Seed field & $M_k (t=0)$ & $\Mach$ & $\MachA (t=0)$ & $\beta (t=0)$ & $\Rm$ & $\Gamma \, (1/t_0)$ & $E_{\rm m}/E_{\rm k}$ & $E_{\rm m}/E_{\rm m0}$ \\
		\hline 
                 \hline	
		 1) Uni & uniform & $M_k(kL/2\pi=0) = N$ & $0.1$ & $3.5 \times 10^{5}$ & $ 2.5 \times 10^{13}$ & $2000$ &  $0.64 \pm 0.03$ & $(3.5 \pm 0.5) \times 10^{-1}$ & $(3.0 \pm 0.3) \times 10^{10}$ \\
		 2) PL  & random & $M_k(2 \le kL/2 \pi \le 20) = N k^{3/2}$ &$0.1$ & $3.5 \times 10^{5}$ & $ 2.5 \times 10^{13}$ & $2000$ &  $0.65 \pm 0.02$ & $(3.3 \pm 0.9) \times 10^{-1}$ & $(2.9 \pm 0.5) \times 10^{10}$ \\
		 3) Par & random & $M_k(19  \le kL/2 \pi \le 21) = N (1 - (k-k_{\rm c})^{2})$ & $0.1$ & $3.5 \times 10^{5}$ & $ 2.5 \times 10^{13}$ & $2000$ &  $0.69 \pm 0.03$ & $(3.3 \pm 0.4) \times 10^{-1}$ & $(2.9 \pm 0.3) \times 10^{10} $ \\
		 \hline 
		 4) UniMA35000 & uniform & $M_k(kL/2\pi=0) = N$ & $0.1$ & $3.5 \times 10^{4}$ & $ 2.5 \times 10^{11}$ &  $2000$ &  $0.67 \pm 0.03$ & $(3.2 \pm 0.6) \times 10^{-1}$ & $(2.8 \pm 0.3) \times 10^{8\phantom{0}}$ \\
		 5) PLMA35000 & random & $M_k(2 \le kL/2 \pi \le 20) = N k^{3/2}$ & $0.1$ & $3.5 \times 10^{4}$ & $ 2.5 \times 10^{11}$ &  $2000$ &  $0.64 \pm 0.08$ & $(2.5 \pm 0.5) \times 10^{-1}$ & $(2.4 \pm 0.3) \times 10^{8\phantom{0}}$ \\
		 6) ParMA35000 & random & $M_k(19 \le kL/2 \pi \le 21) = N (1 - (k-k_{\rm c})^{2})$ & $0.1$ & $3.5 \times 10^{4}$ & $ 2.5 \times 10^{11}$ &  $2000$ &  $0.69 \pm 0.01$ & $(2.7 \pm 0.6) \times 10^{-1}$ & $(2.5 \pm 0.4) \times 10^{8\phantom{0}}$\\
		 \hline 
		 7) UniRm3000 & uniform & $M_k(kL/2\pi=0) = N$ & $0.1$ & $3.5 \times 10^{5}$ & $ 2.5 \times 10^{13}$ & $3000$ &  $0.90 \pm 0.06$ & $(3.1 \pm 0.5) \times 10^{-1}$ &  $(2.7 \pm 0.4) \times 10^{10}$ \\
		 8) PLRm3000 & random & $M_k(2 \le kL/2 \pi \le 20) = N k^{3/2}$ & $0.1$ & $3.5 \times 10^{5}$ & $ 2.5 \times 10^{13}$ &  $3000$ &   $0.92 \pm 0.05$ & $(3.6 \pm 0.7) \times 10^{-1}$ & $(2.9 \pm 0.4) \times 10^{10}$ \\
		 9) ParRm3000 & random &  $M_k(19 \le kL/2 \pi \le 21) = N (1 - (k-k_{\rm c})^{2})$ & $0.1$ & $3.5 \times 10^{5}$ & $ 2.5 \times 10^{13}$ &  $3000$ &  $0.88 \pm 0.02$ & $(3.5 \pm 0.9) \times 10^{-1}$ & $(3.1 \pm 0.5) \times 10^{10}$ \\
		 \hline 
		 10) UniMS10 & uniform & $M_k(kL/2\pi=0) = N$ & $10$ & $3.5 \times 10^{5}$ & $ 2.5 \times 10^{13}$ &  $2000$ &  $0.61 \pm 0.04$ & $(3.6 \pm 0.9) \times 10^{-2}$ & $(4.7 \pm 1.1) \times 10^{13}$ \\
		 11) PLMS10 & random & $M_k(2 \le kL/2 \pi \le 20) = N k^{3/2}$ & $10$ & $3.5 \times 10^{5}$ & $ 2.5 \times 10^{13}$ &  $2000$ &  $0.62 \pm 0.03$ & $(3.6 \pm 0.5) \times 10^{-2}$ & $(4.4 \pm 0.7) \times 10^{13}$ \\
		 12) ParMS10 & random & $M_k(19 \le kL/2 \pi \le 21) = N (1 - (k-k_{\rm c})^{2})$ & $10$ & $3.5 \times 10^{5}$ & $ 2.5 \times 10^{13}$ &  $2000$ &  $0.59 \pm 0.06$ & $(4.7 \pm 0.7) \times 10^{-2}$ & $(6.0 \pm 0.9) \times 10^{13}$ \\
		 \hline 
		 13) UniRm100 & uniform & $M_k(kL/2\pi=0) = N$ & $0.1$ & $3.5 \times 10^{5}$ & $ 2.5 \times 10^{13}$ &  $100$ &  $4.74 \pm 0.01$ & $(6.6 \pm 0.9) \times 10^{-11}$ & $9.5 \pm 1.0 $ \\
		 14) PLRm100 & random & $M_k(2 \le kL/2 \pi \le 20) = N k^{3/2}$ & $0.1$ & $3.5 \times 10^{5}$ & $ 2.5 \times 10^{13}$ &  $100$ &  decaying & $\sim 0$ & $\sim 0$ \\
		 15) ParRm100 & random & $M_k(19 \le kL/2 \pi \le 21) = N (1 - (k-k_{\rm c})^{2})$ & $0.1$ & $3.5 \times 10^{5}$ & $ 2.5 \times 10^{13}$ &  $100$ &  decaying & $\sim 0$ & $\sim 0$ \\
		 \hline 		 
		\hline
	\end{tabular}
\end{table*}

For the seed magnetic field, we use three different setups with the same total magnetic field strength, 
\begin{itemize}
\item $B_0\ne0$, Uni: non-zero seed mean field, which is modelled as a uniform magnetic field throughout the domain along the $z$ direction, i.e, $M_k(kL/2\pi=0) = N$, where
$N$ is a constant.
\item $B_0=0$, PL: small-scale random field with zero mean field and power at a range of scales, which is modelled by generating
random magnetic fields with a power-law magnetic spectrum (slope chosen to be $3/2$, motivated by the Kazantsev spectrum, \cite{Kazantsev1968}), i.e., $M_k(2 \le kL/2 \pi \le 20) = N k^{3/2}$, and
\item $B_0=0$, Par: small-scale random field with zero mean field $B_0=0$ and localised magnetic power at small scales, which is modelled by generating
random magnetic fields with a parabolic magnetic spectrum, i.e., $M_k(19 \le kL/2 \pi \le 21) = N (1 - (k-k_{\rm c})^{2})$ with a peak at the critical wave number $k_{\rm c} = 20 (2 \pi /L)$.
\end{itemize}
The constant $N$ is chosen such that required seed field strength (or initial Alfv\'en Mach number) is achieved. 
\Fig{inicond}(a) shows the computed magnetic power spectrum for random seed magnetic fields and for $B_0=0$, PL case the slope of power-law magnetic spectrum is $1.53\pm0.06$, 
very close to the input value of $1.5$. The probability distribution function (PDF) of the normalised magnetic field $z$ component $b_z/\brms$ for all 
three seed magnetic fields in shown in \Fig{inicond}b. The PDF for random seed fields is very close to a Gaussian distribution with zero mean
and one-third standard deviation. Both random fields have a different coherence length $\lb$, defined by (in term of the numerical domain size, $L$)
\begin{equation}
\lb/L = \frac{\int_0^{\infty} k^{-1} M_k \, \dd k}{\int_0^{\infty} M_k \, \dd k}.
\label{lb}
\end{equation}
For $B_0=0$, PL, $\lb/L =0.08$, and for $B_0=0$, Par, $\lb \simeq 1/20 = 0.05$.
The strength of the seed magnetic field $\Brms (t=0)$ is chosen to reach Alfv\'en Mach number, $\MachA (t=0) = (4\pi\rho \, (t=0))^{1/2} \urms / \Brms (t=0) = 3.5 \times 10^{5}$ and
$3.5 \times 10^{4}$. 

\Tab{param} gives the parameters of our simulations runs. 
Runs 1 (Uni), 2 (PL), and 3 (Par) are our base runs with $B_0\ne0$, Uni, $B_0=0$, PL, and  $B_0=0$, Par seed magnetic fields and 
other parameters as $\Mach = 0.1, \MachA (t=0) = 3.5\times10^{5}, \beta (t=0) = 2.5 \times 10^{13}, \Re = 2000,$ and $\Rm = 2000$.
We then vary each parameter for all three seed fields and this change from the base runs is also reflected in the simulation name. For example,
when the $\Rm$ is changed to $3000$, the simulation name is UniRm3000, PLRm3000, and ParRm3000 for $B_0\ne0$, Uni, $B_0=0$, PL, and  $B_0=0$, Par seed magnetic fields.
We run each simulation until the magnetic fields reaches a statistically steady state, 
which for this set of parameters is approximately $100 t_0$ for dynamo runs and $ 20 t_0$ when $\Rm < \Rmc$ with non-zero mean seed field, where $t_0$ is the eddy turnover time.

Throughout the paper we use non-dimensional units to describe physical quantities: lengths 
in units of the numerical domain length $L$, time in units of the eddy turnover time $t_0$, 
speeds in units of the sound speed $\cs$ (or Mach number, $\Mach$), density in terms of the initial density $\rho \, (t=0)$, the magnetic field in
units of $(4 \pi \rho \, (t=0) \cs^2)^{1/2}$ (or Alfv\'en Mach number, $\MachA$), 
and all the diffusivities in units of $\cs L$. 
The actual numbers can be scaled depending on the physical situation (chosen density and temperature of the turbulent medium).
For example, for the warm phase (which has the maximum volume filling fraction) of the ISM in spiral galaxies
(temperature $\approx 10^{4} \K$, Hydrogen number density $\approx 10^{-1} \cm^{-3}$), $L\simeq100\pc$, $t_0\simeq0.5 \times 10^{7} \yr$, 
$\cs\simeq10 \kms$, $\rho \, (t=0) \simeq 10^{-25} \g \cm^{-3}$,  $(4 \pi \rho \, (t=0) \cs^2)^{1/2} \simeq 10^{-10} \G$, 
and $\cs L \simeq 3 \times 10^{26} \cm^2 \s^{-1}$. For results with non-zero seed mean field ($B_0\ne0$, Uni), we subtract the mean seed field $B_0$ from the dynamo-generated field $B$ to obtain the small-scale random field $b$ with mean zero, which is the focus of this study. However, since the mean-field energy is orders of magnitude smaller than the energy in the dynamo-generated small-scale field, whether or not we subtract the mean field does affect any of our results.

\section{Results} \label{res}

\subsection{Time evolution of dynamo amplification}

\begin{figure*}
\includegraphics[width=\columnwidth]{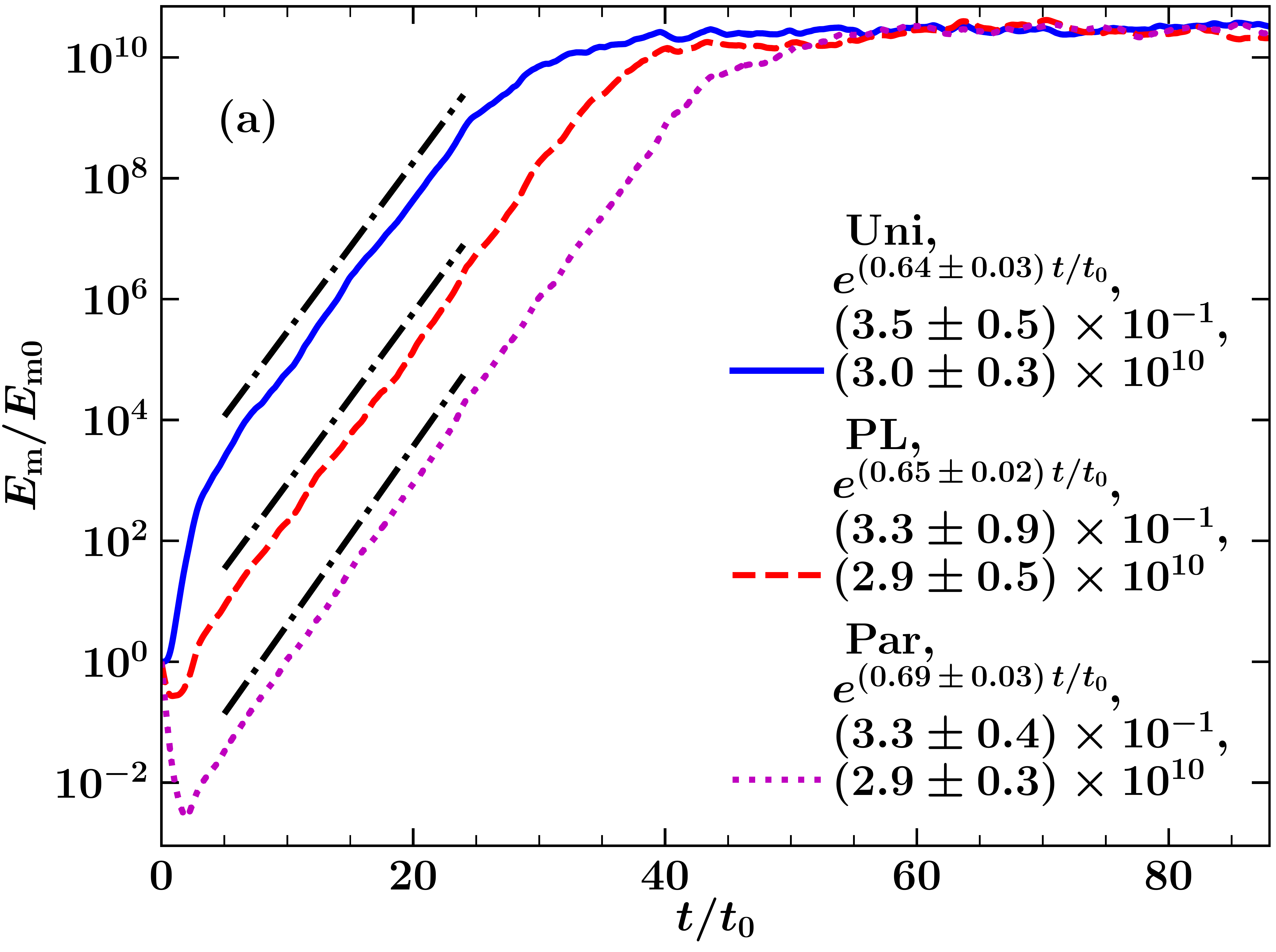}  \hspace{0.5cm}
\includegraphics[width=\columnwidth]{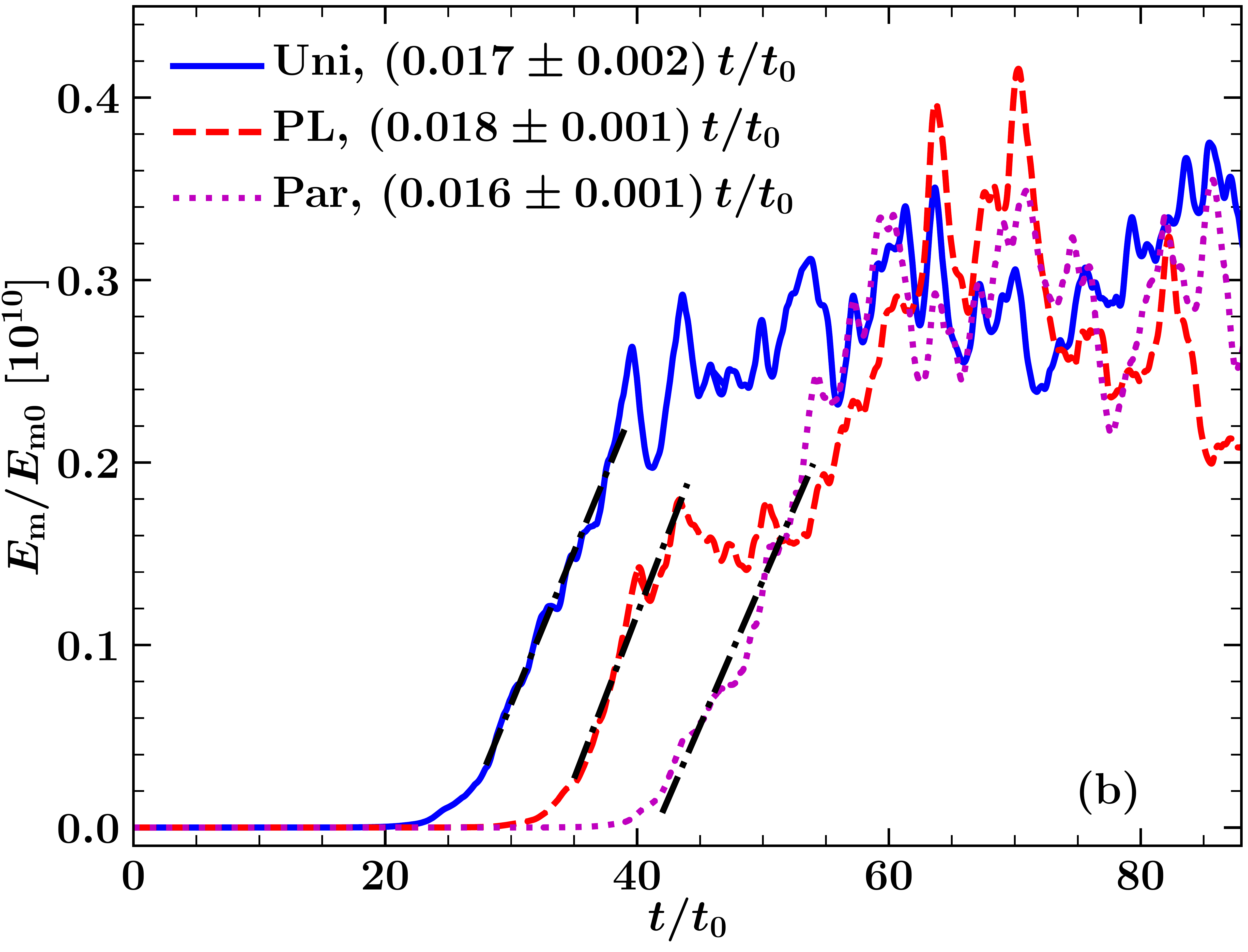}
\caption{(a) Time evolution of magnetic energy $E_{\rm m}/E_{\rm m0}$ (where $E_{\rm m0}$ is the magnetic energy at $t=0$ and $t_0$ is the eddy turnover time)
for models: Uni (solid, blue), PL (dashed, red) and Par (dotted, magneta). 
For $t/t_0 \lesssim 3$ (transient phase), the uniform seed magnetic field (Uni case) grows very quickly whereas the random fields decays. 
Then, once the magnetic field becomes the eigenfunction of the induction equation, all three seed fields grow exponentially (dash-dotted, black) with very similar growth rates (shown with associated errors in the legend). Finally, the field saturates for all of them ($t/t_0 \gtrsim 50$) due to the back-reaction of the magnetic field by the Lorentz force. In the legend, after the simulation name and
the fitted function to quantify growth rate, we provide the final saturated level in terms
of $E_{\rm m}/E_{\rm k}$ and $E_{\rm m}/E_{\rm m0}$.
The saturated level magnetic level is also the same for all three seed fields, 
$E_{\rm m}/E_{\rm k} \simeq 3.4\times10^{-1}$ and $E_{\rm m}/E_{\rm m0} \simeq 3\times10^{10}$. (b) Same as (a) but with a linear scale on the $y$-axis to highlight the
transition region from the kinematic to saturated stage. In the transition region, the magnetic field grows linearly with time (as shown by the dash-dotted, black lines), and we find that this linear growth rate is roughly the same in all three seed-field cases (fit parameters and uncertainties are shown in the legend).}
\label{ts}
\end{figure*}

First, we show the results for our standard set of runs: Uni, PL, and Par. \Fig{ts}(a) shows the evolution of the magnetic field
energy for all three seed-field cases. After the initial transient phase ($t/t_0 \ge 3$), the magnetic field for all three cases, first grows
exponential (kinematic stage) with very similar growth rate ($\approx 0.66/t_0$) 
and then achieves a statistically steady state (saturated stage) with the same saturated level ($E_{\rm m}/E_{\rm k} \simeq 3.4\times10^{-1}$ and $E_{\rm m}/E_{\rm m0} \simeq 3\times10^{10}$). 
Thus, the structure of the seed magnetic field does not affect the growth rate (column 8 in \Tab{param}) and the saturated level (columns 9 and 10 in \Tab{param}) of magnetic energy in the small-scale turbulent dynamo.

In the kinematic stage of the small-scale dynamo, the back-reaction of the magnetic field on the velocity field is negligible because the magnetic field is very weak. Thus, the exponential amplification in the kinematic stage can be achieved by a linear operator on the seed field. When the magnetic field becomes strong enough to back react on the flow, the non-linearity kicks in and the exponential growth rate slows down (shown in \Fig{ts}(a)). Then the magnetic energy evolves linearly with time as shown in \Fig{ts}(b) \citep[also seen by][]{ChoEA2009}. The growth rate, even in this transitional phase, is roughly the same for all three seed-field cases, as quantified by a linear fit (dash-dotted black lines), with the fitted slopes listed in the figure legend. The magnetic field finally reaches a statistically saturated state, where any memory of the seed field is expected to be lost.

\subsection{Magnetic field spectra and coherence lengths}

\begin{figure*}
\includegraphics[width=\columnwidth]{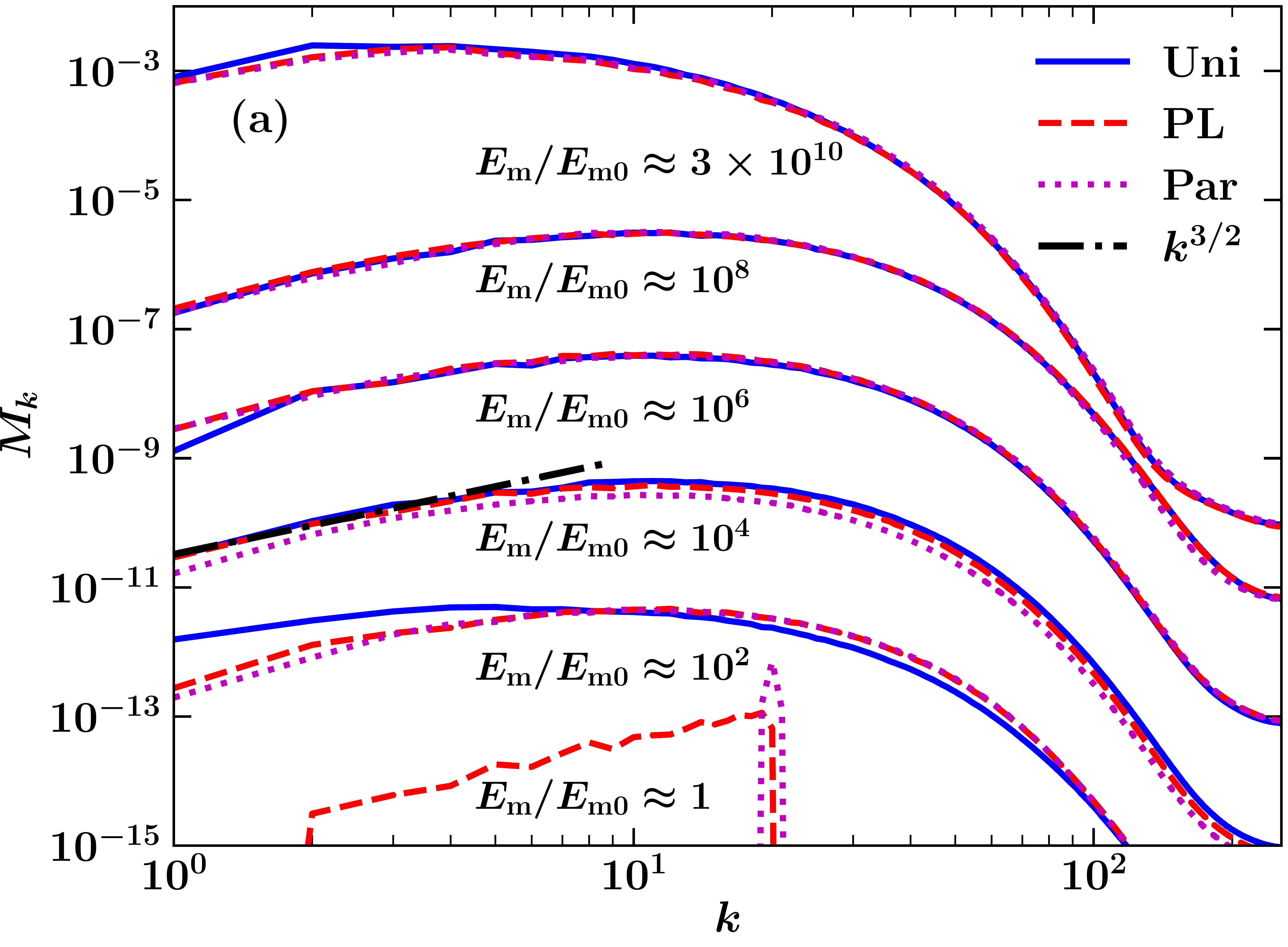} \hspace{0.5cm}
\includegraphics[width=\columnwidth]{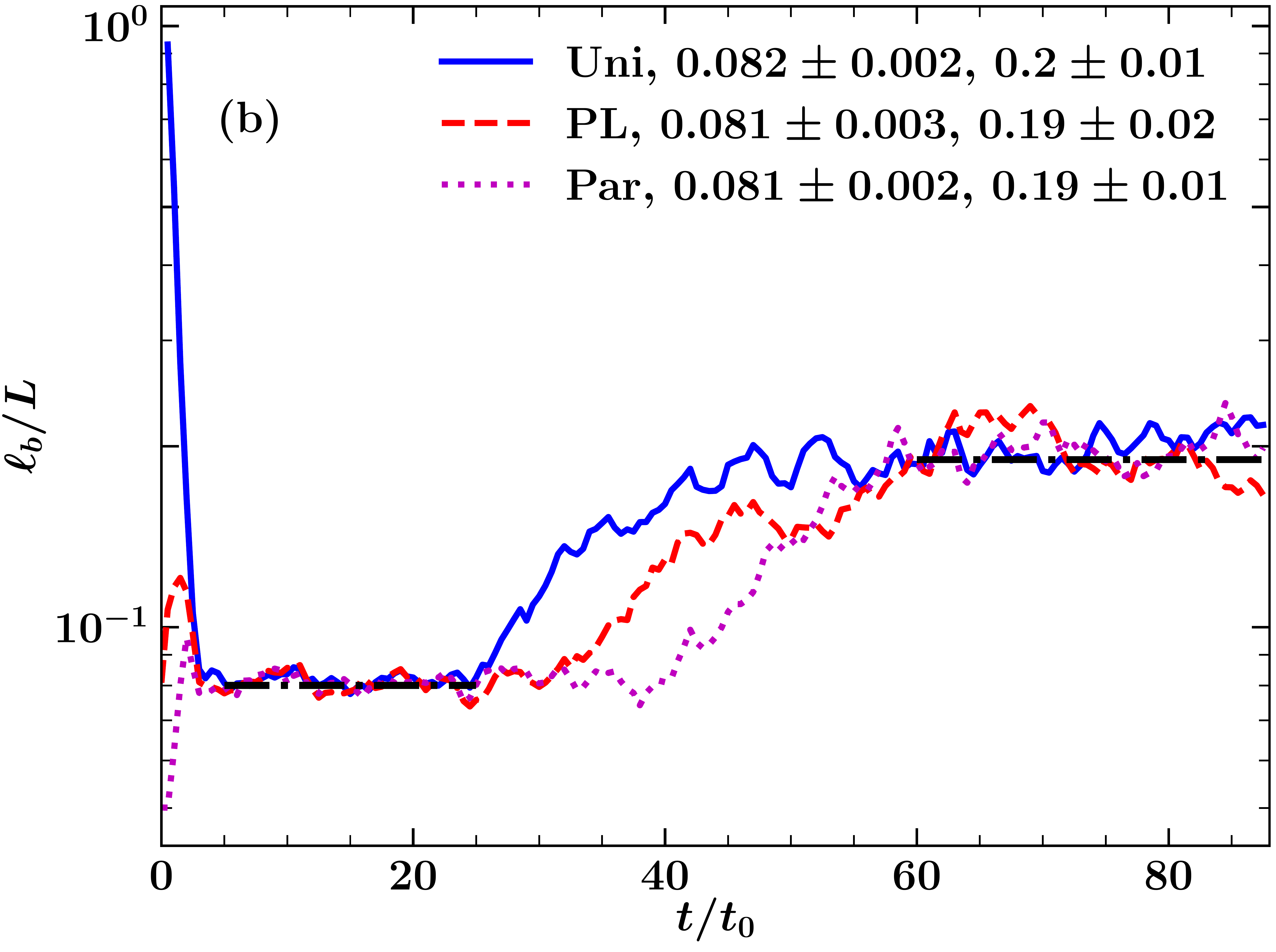}
\caption{(a) The shell-averaged magnetic power spectrum for models: Uni (solid, blue), PL (dashed, red), and Par (dotted, magenta)
at various stages of their magnetic field evolution ($E_{\rm m}/E_{\rm m0} \approx 1, 10^2, 10^4, 10^6, 10^8, 3 \times 10^{10})$.  
Initially, the seed field ($E_{\rm m}/E_{\rm m0} \approx 1$), even for the localised parabolic field (dashed, magenta), 
quickly evolves to have power at a range of wave numbers  (see \Fig{ts}(a) for different times at which  
the amplification of $E_{\rm m}/E_{\rm m0} \approx 10^{2}$ is achieved by different seed fields). The
spectrum for the seed field for Uni shows power on scales larger than the random field cases (PL and Par).
But once the fields reach the kinematic stage, the power spectrum for all three seed field cases is very similar
and follows the Kazantsev $k^{3/2}$ spectrum at lower wave numbers ($1 \le kL/2 \pi \le 10$). The field then saturates
with its power spectrum becoming flatter on larger scales. The power spectrum for all three seed-field cases is very similar
in the kinematic and saturated stages.
(b) The coherence length $\lb/L$ calculated using \Eq{lb} for all three seed-field cases, Uni (solid, blue), PL (dashed, red), and Par (dotted, magenta),
as a function of time, $t/t_0$ (refer to \Fig{ts}(a) for converting the time value on the $x$-axis to the amplification factor,  $E_{\rm m}/E_{\rm m0}$). 
In the legend, the average of the coherence length and one standard deviation variation (reported as error) 
in the kinematic ($10 \le t/t_0 \le 20$) and saturated ($t/t_0 \ge 60$) stages are given. 
Even though all three magnetic fields have very different correlation length in the beginning, they end up having very similar correlation lengths
in the kinematic and saturated stages. The coherence length is higher in the saturated stage compared to the kinematic stage, because the field develops more large-scale power once it saturates.
}
\label{spec}
\end{figure*}

Now we take a look at the spectral properties of the magnetic fields in the three runs, as they evolve from the seed to the kinematic stage, and finally to the saturated stage. \Fig{spec}(a) shows the shell-averaged magnetic power spectrum for all three seed field cases (Uni, PL, and Par) at various times in their magnetic field evolution. 
In the seed-field stage, all three cases have a very different spectral shape. The power for the uniform seed-field case (Uni) only exists at $k=0$. For the random field cases (PL and Par), the power is spread over a range of wavenumbers ($2 \le kL/2 \pi \le 20$) with the power distributed as a power law (slope $3/2$; PL), or localised in wavenumber space with a parabolic function peaking at $kL/2 \pi = 20$ (Par).
The field for all three cases spreads very quickly in wavenumber space and occupies power on a range of scales. In the initial transient phase, before the magnetic energy enters the exponential growth phase, $E_{\rm m}/E_{\rm m0} \approx 10^{2}$ in \Fig{ts}(a) for Uni, the spectra show higher power on larger scales in Uni than in PL and Par. This is due to additional tangling of the mean field in Uni. Then the magnetic field in the kinematic stage follows a Kazantsev $k^{3/2}$ spectrum \citep{Kazantsev1968} for low wavenumbers ($1 \le kL/2 \pi \le 10$) and the spectra become flatter as the magnetic field saturates. For all three seed-field cases, the magnetic field spectrum is very similar in their respective kinematic and saturated stages.

Based on the magnetic field spectrum, we also calculate the coherence length (normalised to the numerical domain size, $L$), 
$\lb/L$, of the magnetic fields for all three cases using \Eq{lb}. The time evolution of $\lb/L$ is shown in \Fig{spec}(b). 
For $10 \le t/t_0 \le 20$ (kinematic stage), $\langle \lb/L \rangle$, is approximately equal to  $0.082 \pm 0.002$, $0.081 \pm 0.003,$ and $0.081 \pm 0.002$ for Uni, PL, and Par, respectively (where $\langle \cdots \rangle$ denotes average over time and the reported error is one standard deviation). The corresponding values in the saturated stages ($t/t_0 \ge 60$) are $\langle \lb/L \rangle = 0.20 \pm 0.01$, $0.19 \pm 0.02$, and $0.19 \pm 0.01$. The magnetic field coherence length is higher in the saturated stages as compared to the kinematic stage. However, the coherence length, which is different to begin with (by construction), eventually becomes very similar in the kinematic stage and saturated stages for all three seed-field cases. 

\subsection{Magnetic field morphology}

\begin{figure*}
\includegraphics[width=2\columnwidth]{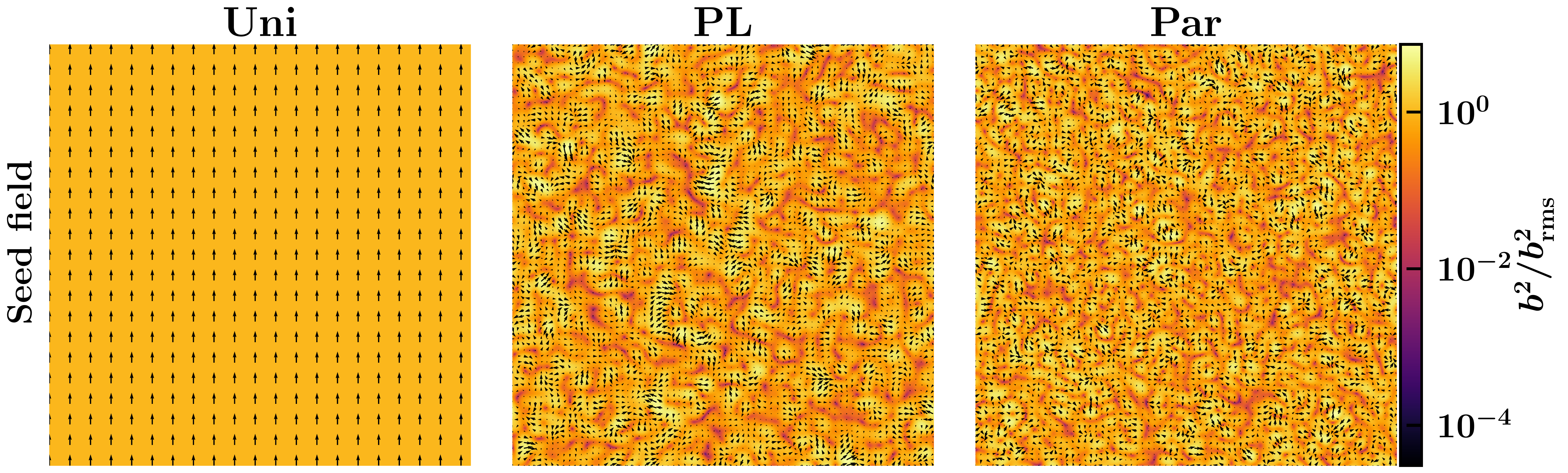} \\
\includegraphics[width=2\columnwidth]{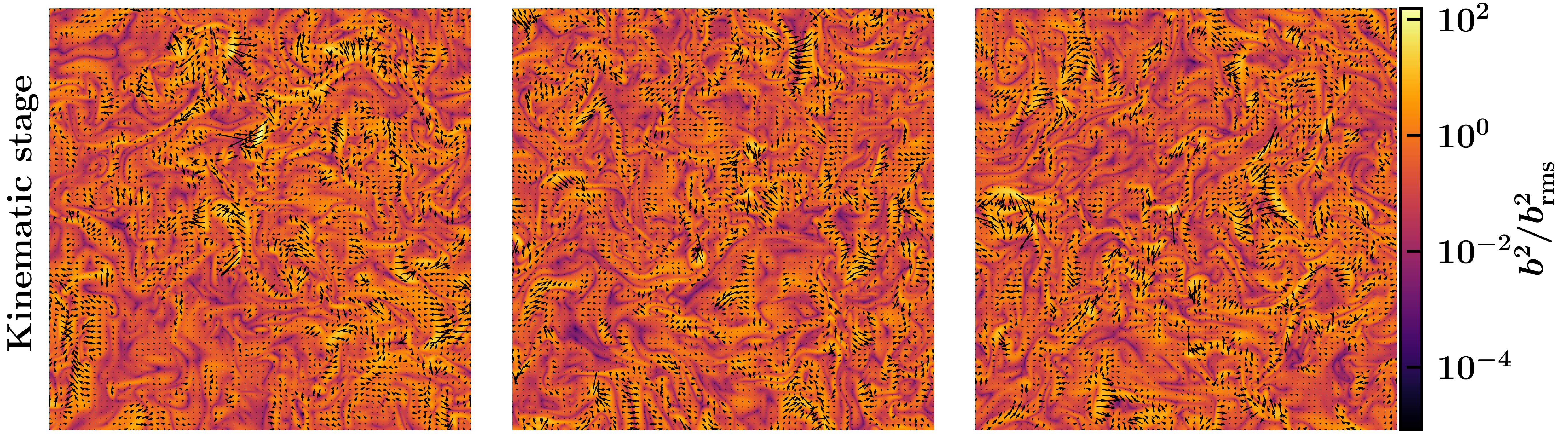}\\
\includegraphics[width=2\columnwidth]{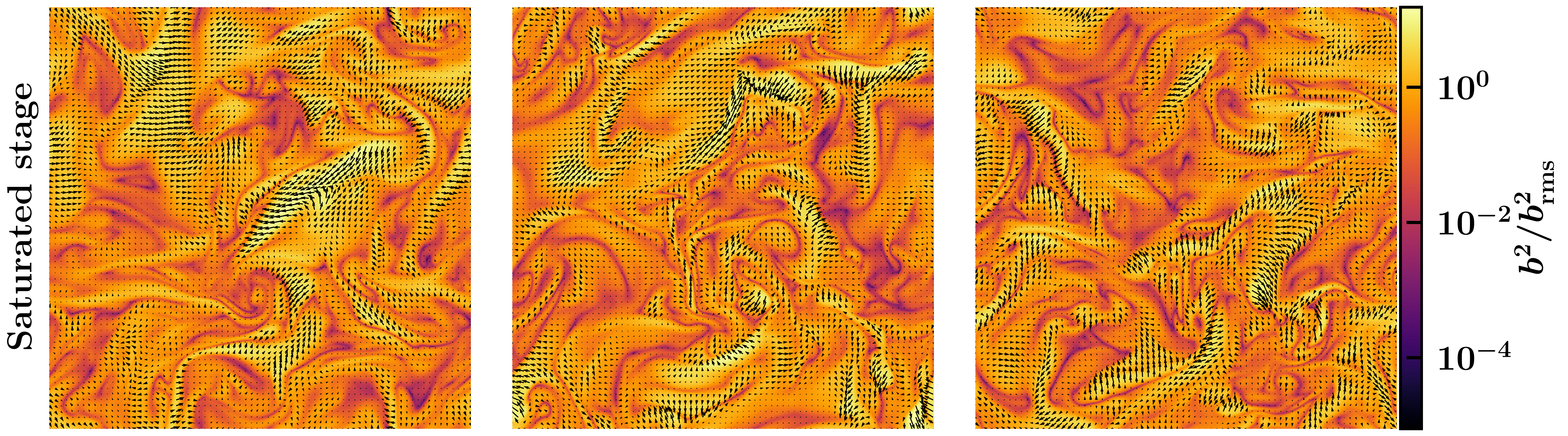} 
\caption{Two-dimensional slices in the $xz$ plane with vectors representing $(b_x/\brms,b_z/\brms)$ and the colour showing the magnitude of the normalised
 magnetic energy $b^2/\brms^2$ for all three seed fields: Uni (first column), PL (second column), and Par (third column) in the seed (first row),
 kinematic  (second row), and saturated  (third row) stages. The seed magnetic field is uniform for the Uni case (first row, first column) and is
 random for the PL (first row, second column) and Par (first row, third column) cases with the PL seed field visually showing structures slightly larger than
 that for the Par field. For the kinematic and saturated stages, the fields looks very similar for all three cases. The magnetic fields in the saturated stages
 show structures larger than that in the kinematic stages.}
\label{2dplots}
\end{figure*}

After confirming that the time evolution and the magnetic field spectra (and coherence lengths) 
are not affected by the structure of seed magnetic fields, we explore the effect 
of the seed field on the structure of the small-scale dynamo generated magnetic fields. In \Fig{2dplots}, we first show two-dimensional slices with vectors representing $(b_x/\brms, b_z/\brms)$, and with the colour showing the normalised magnetic energy $b^2/\brms^2$ for all three seed-field cases (Uni, PL, and Par) and
for all three stages of the small-scale dynamo (seed, kinematic stage, and saturated stage). The seed fields (first
row in \Fig{2dplots}) show a uniform field for the Uni case and random fields for PL and Par cases, as expected, with structures smaller in size
for the Par case. In the kinematic stage (second row in \Fig{2dplots}) and saturated stage (third row in \Fig{2dplots}), the magnetic field structures are statistically very similar 
for all three cases. The structures in the kinematic stage are bigger in size (visually) than the saturated stage \citep[consistent with the discussion in][]{SetaEA2020}.
This also agrees with the higher coherence length in the saturated stage in comparison to the kinematic stage as previously shown in \Fig{spec}(b).

\subsection{Probability density functions (PDFs) of the magnetic field}

\begin{figure*}
\includegraphics[width=\columnwidth]{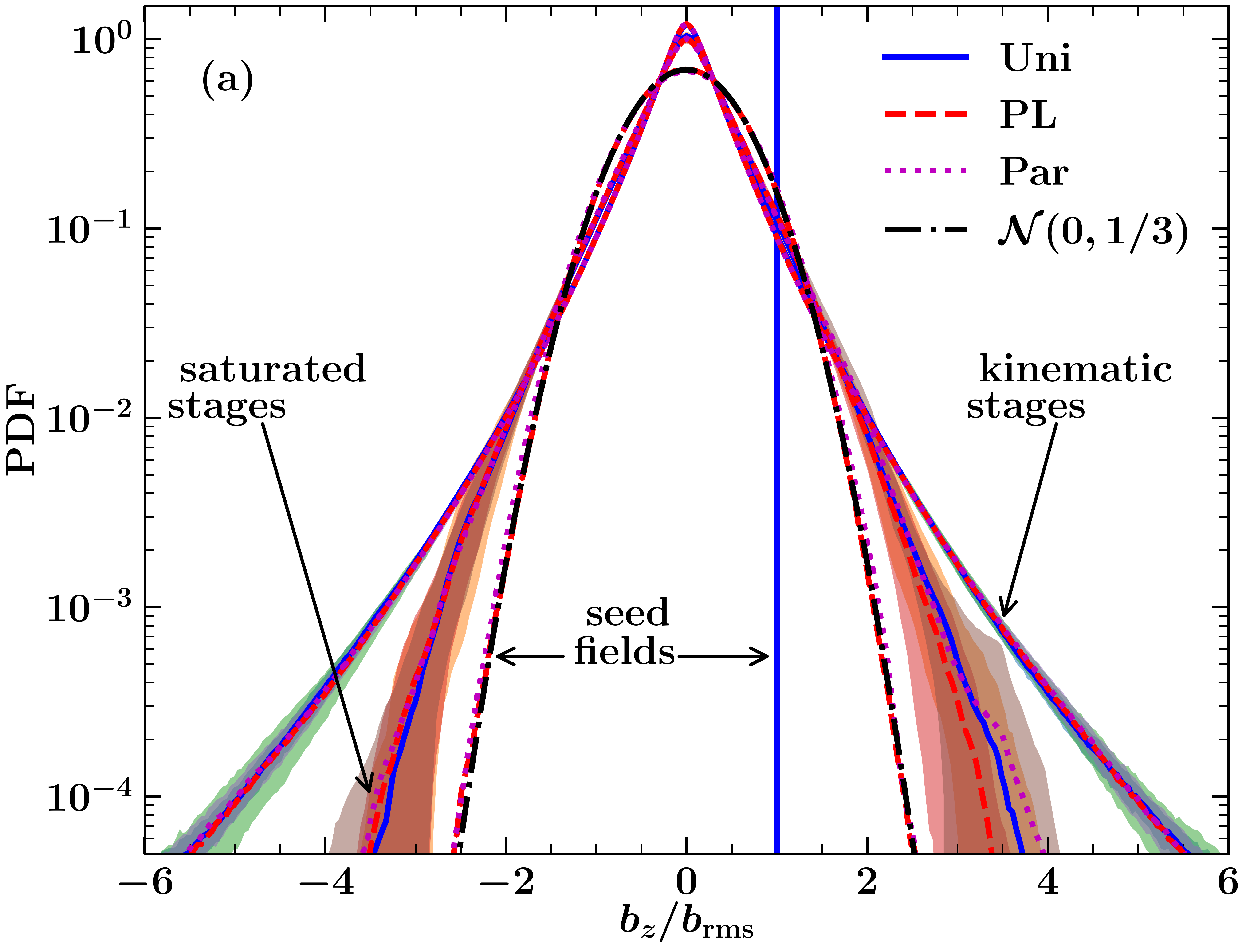} \hspace{0.5cm}
\includegraphics[width=\columnwidth]{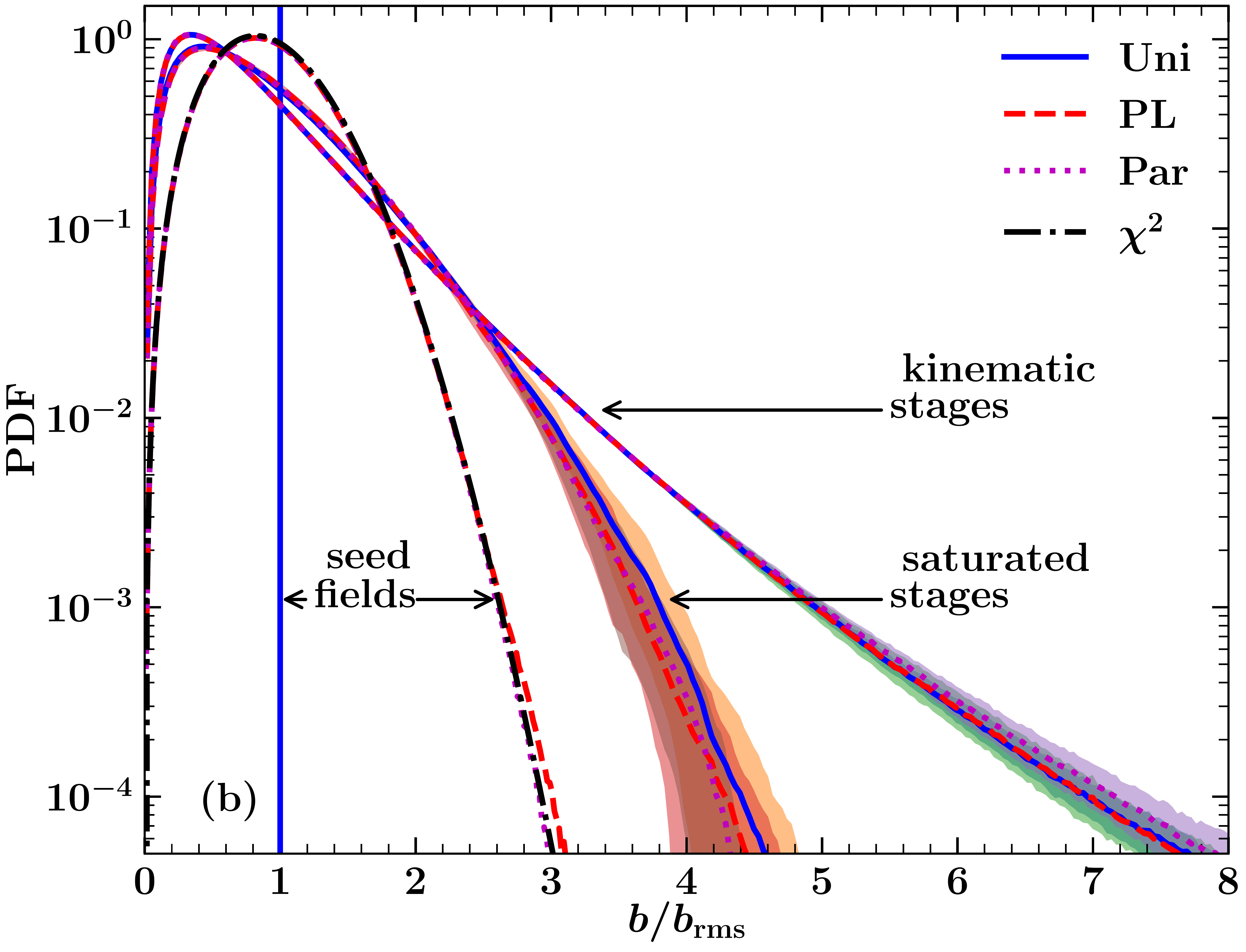}
\caption{(a) The probability density function (PDF) of the normalised $z$ component $b_z/\brms$ of the magnetic fields for models with all 
three seed-field cases: Uni (solid, blue), PL (dashed, red), and Par (dotted, magenta), 
in the seed, kinematic, and saturated stages. The shaded regions (shown in different colours) around the PDFs indicate the 1-sigma variations when averaged over $10$ eddy turnover times ($t_0$) in the
kinematic ($10 \le t/t_0 \le 20$) and saturated ($80 \le t/t_0 \le 90$) stages. The PDFs for all three seed-field cases are very similar in both the 
kinematic and saturated stages, with the field being more intermittent (or equivalently non-Gaussian) in the saturated stage 
(shown by heavy tails in the kinematic stage for values $|b_z|/\brms \gtrsim 4$). 
(b) The PDF of $b/\brms$ for magnetic fields in all three-seed field cases in the seed, kinematic and saturated stages,
with the shaded region showing the 1-sigma variations when averaged over $10 t_0$ for the kinematic ($10 \le t/t_0 \le 20$) 
and saturated ($75 \le t/t_0 \le 85$) stages. The PDFs in the seed-field stage follows a $\chi^2$ (dashed, black) distribution 
for the PL and Par simulations. The PDFs for all three seed fields are very similar in the kinematic and saturated stages, with the field 
being less intermittent in the saturated stage where the non-linearity due to the Lorentz force truncates $b/\brms$ approximately around $4.5$.}
\label{pdfs}
\end{figure*}

In \Fig{pdfs}, we show the PDFs for the $z$ component of the magnetic field $b_z/\brms$ and 
the magnetic field strength $b/\brms$ for all three seed magnetic fields for the seed, kinematic, and saturated stages. 
For both figures in the kinematic and saturated stage, the shaded region around the lines shows the 1-sigma variations
around the mean value. The PDFs for PL and Par follow a Gaussian distribution for $b_z/\brms$ and a $\chi^2$ distribution (the sum of the square of
independent normal distributions) for $b/\brms$. In the kinematic stage,
the magnetic field is more intermittent (even after taking into account the statistical fluctuations), as shown by heavier tails for the higher values of $b_z/\brms$ and $b/\brms$ in \Fig{pdfs}(a) and \Fig{pdfs}(b), respectively. Thus, the saturated stage is more Gaussian or volume filling than the kinematic stage (this has also been shown before by a variety of approaches in \cite{SetaEA2020}). However, the PDFs are statistically similar for all three seed magnetic fields (so even the statistical moments of the small-scale random magnetic fields, including higher-order ones, will also be similar for all three seed-field cases). Thus, based on \Fig{2dplots} and \Fig{pdfs}, we conclude that the configuration of the seed magnetic field does not affect the statistical structure of the field later on, in both
the kinematic and saturated stages of the small-scale dynamo.

\subsection{Dependence on the seed magnetic field strength, the magnetic Reynolds number, and the sonic Mach number}

\begin{figure*}
\includegraphics[width=0.69\columnwidth]{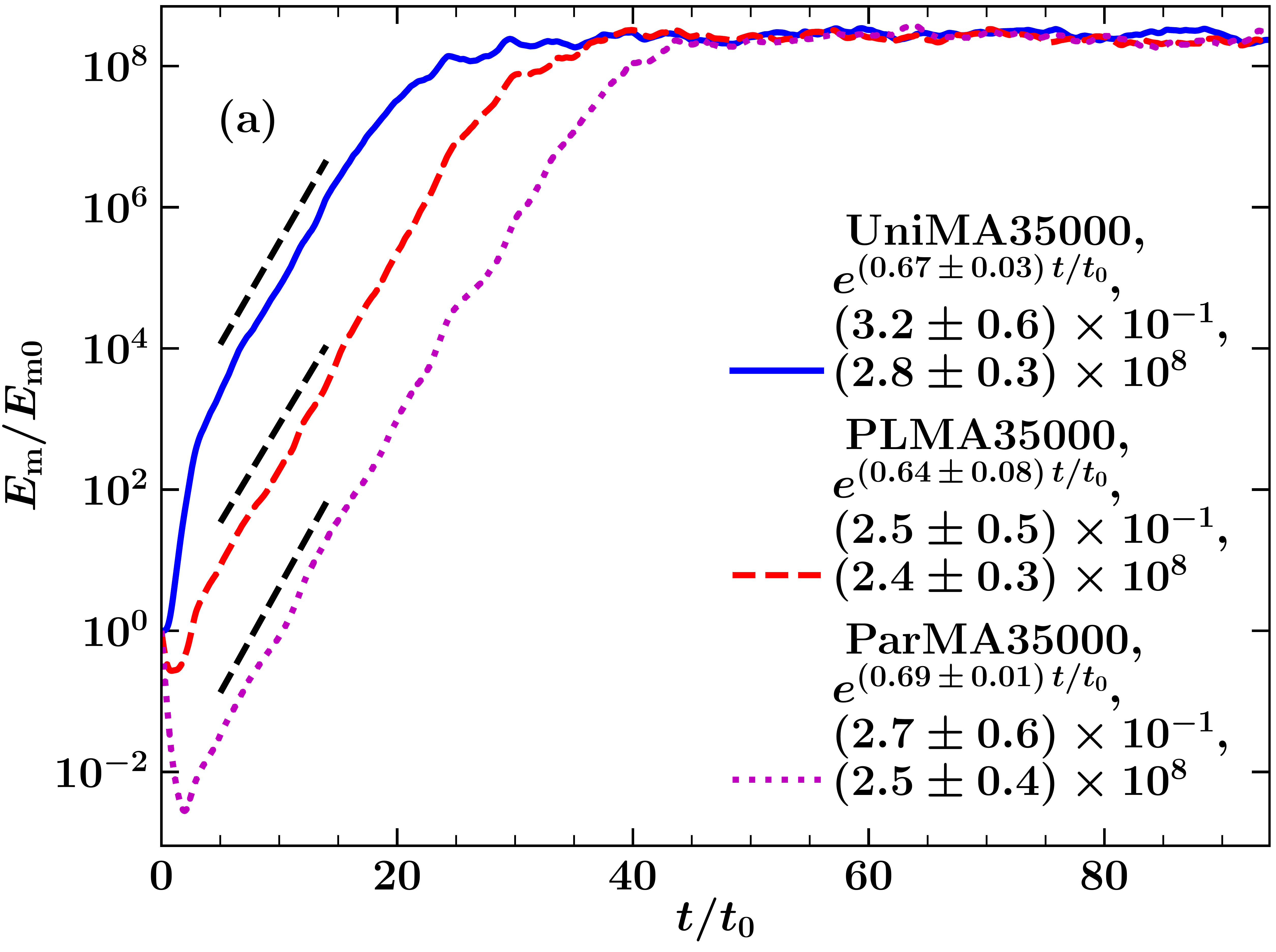}
\includegraphics[width=0.69\columnwidth]{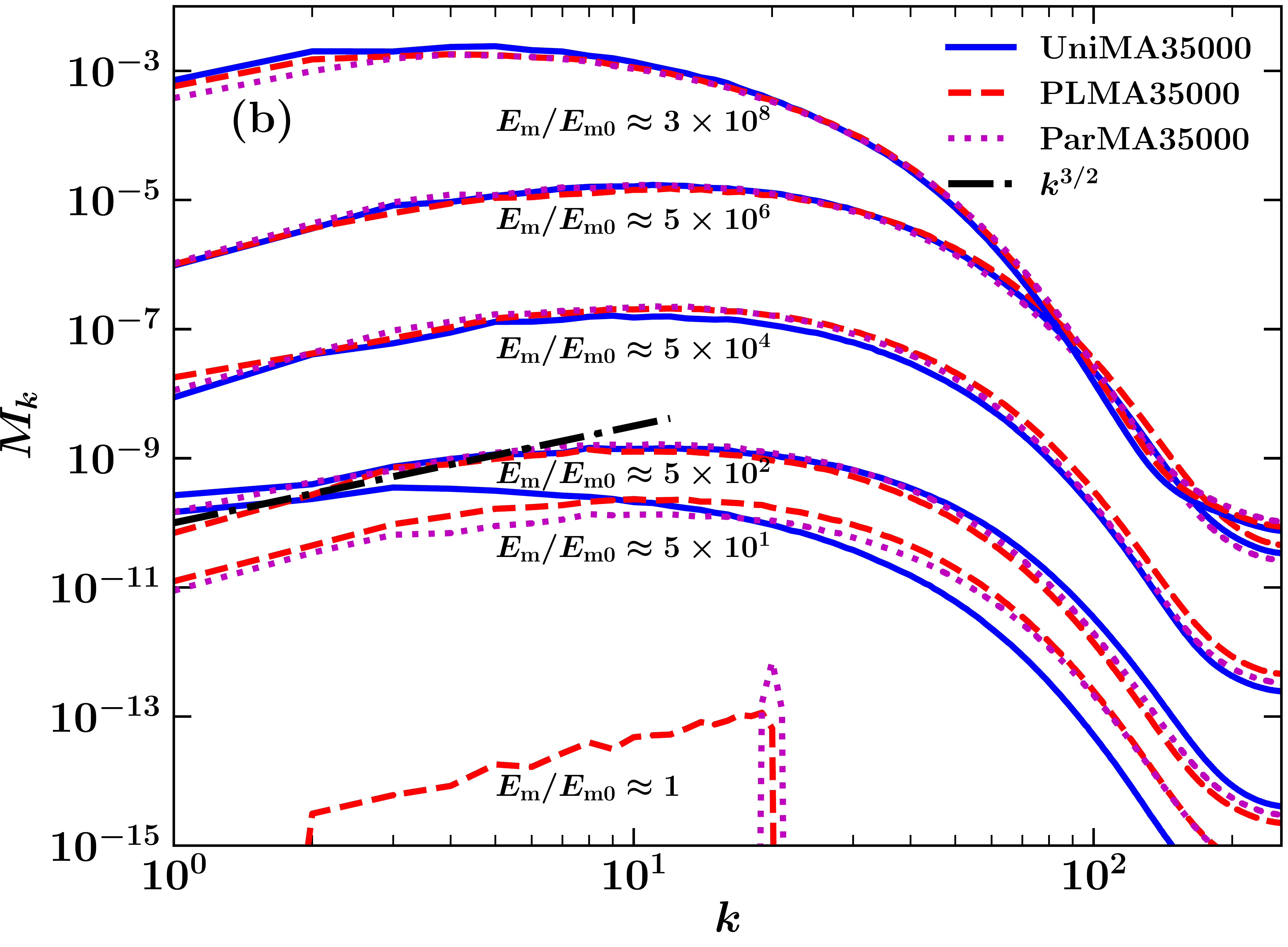}
\includegraphics[width=0.69\columnwidth]{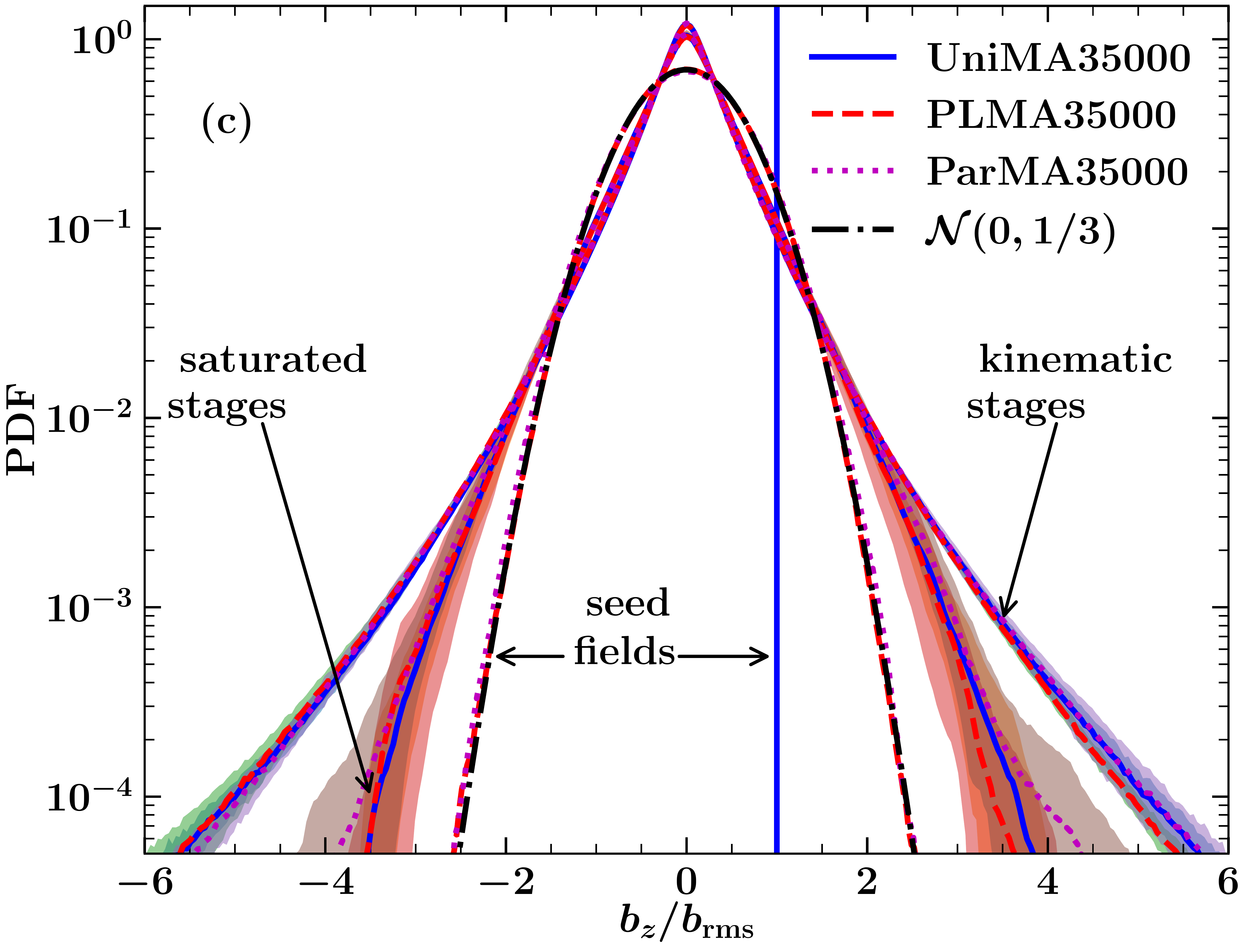}
\caption{Time evolution of magnetic energy (left), magnetic spectra (middle), and PDFs of the $z$ component of the magnetic field (right). These panels are the same as in \Fig{ts}(a), \Fig{spec}(a), and \Fig{pdfs}(a), but here we show them for a different seed magnetic field strength corresponding to $\MachA=3.5\times10^{4}$ (models UniMA35000, PLMA35000, and ParMA35000; see \Tab{param}). We find that the structure of the seed magnetic field does not significantly alter the dynamo growth rate, saturated level, magnetic power spectrum and 
PDF of the field in the kinematic and final saturated stages, and this conclusion is furthermore independent of the choice of the seed strength of the magnetic field \citep[as long as the seed magnetic field is sufficiently weak to allow for dynamo amplification; see e.g.,][for the different regimes of MHD turbulence]{Federrath2016}.}
\label{b0}
\end{figure*}

\begin{figure*}
\includegraphics[width=0.69\columnwidth]{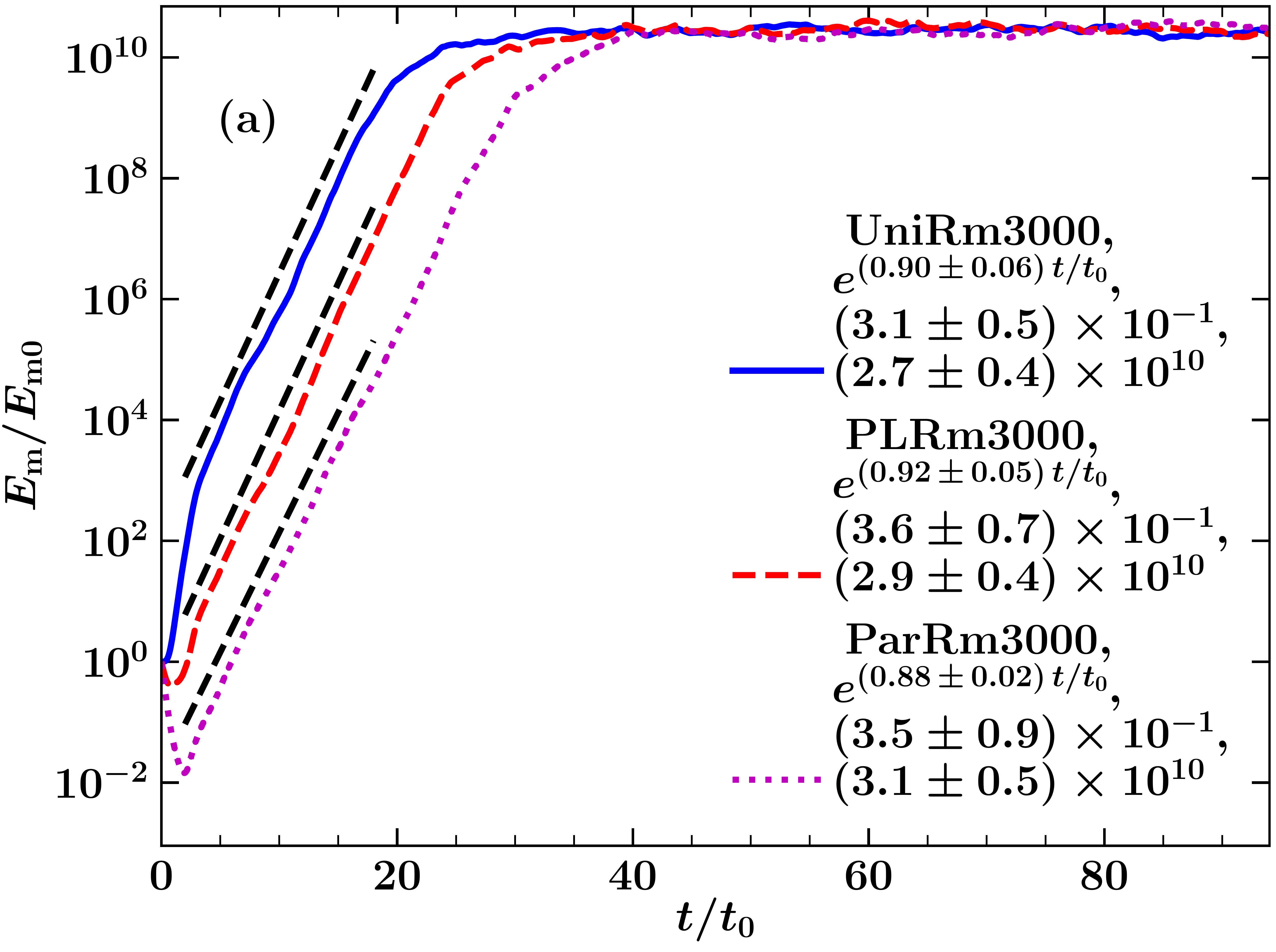}
\includegraphics[width=0.69\columnwidth]{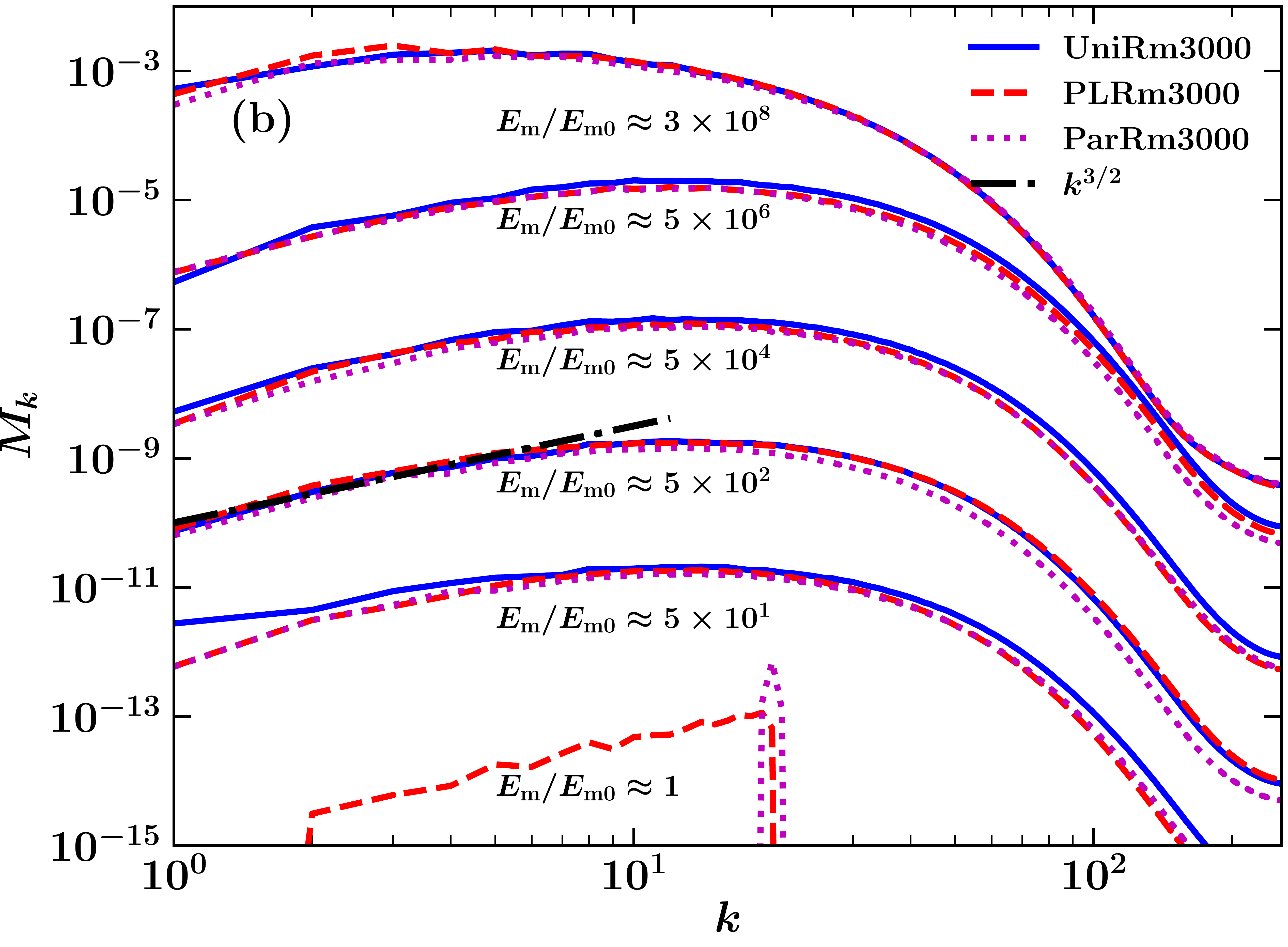}
\includegraphics[width=0.69\columnwidth]{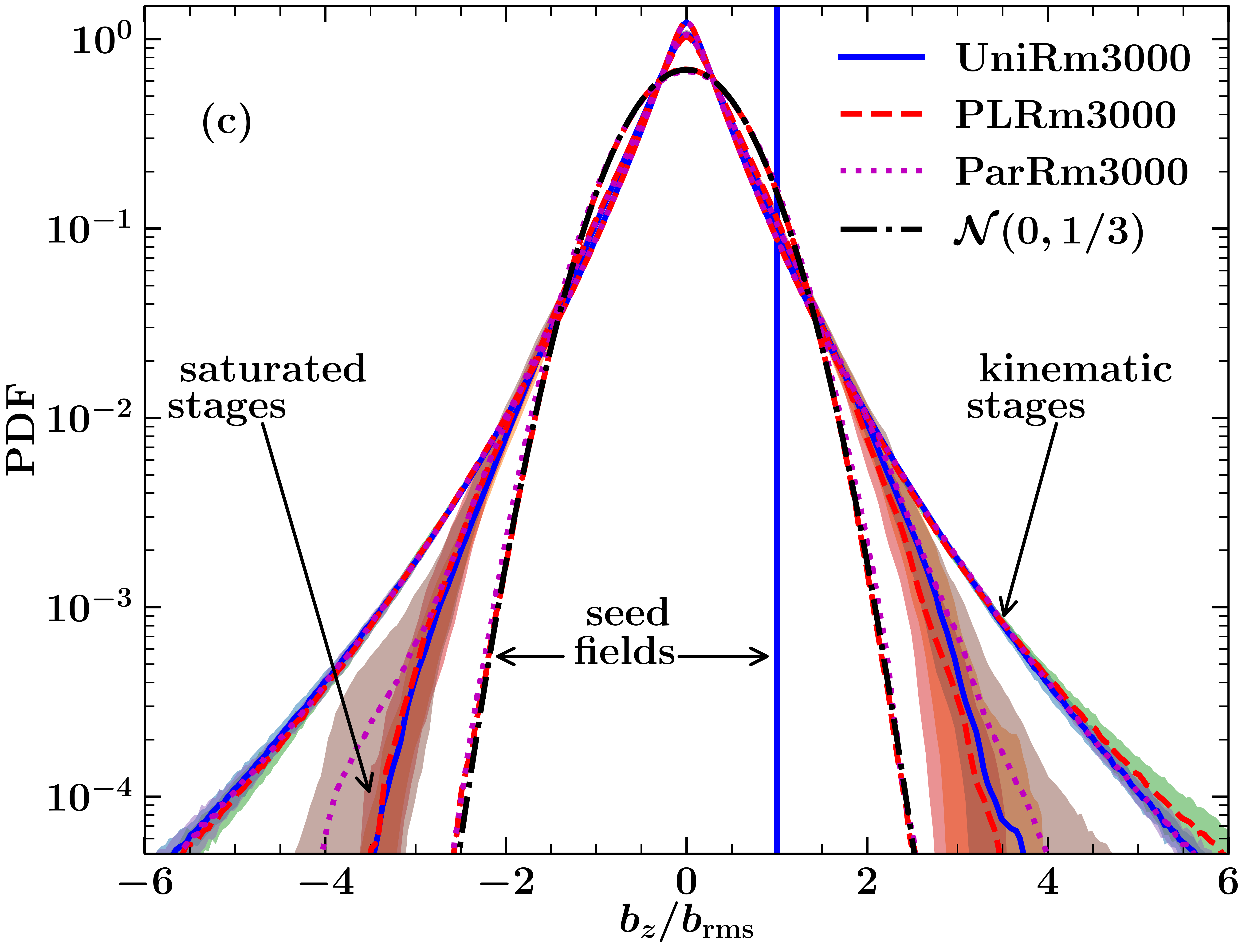}
\caption{Same as \Fig{b0}, but for $\Rm=3000$ (models UniRm3000, PLRm3000, and ParRm3000).}
\label{rm}
\end{figure*}

\begin{figure*}
\includegraphics[width=0.69\columnwidth]{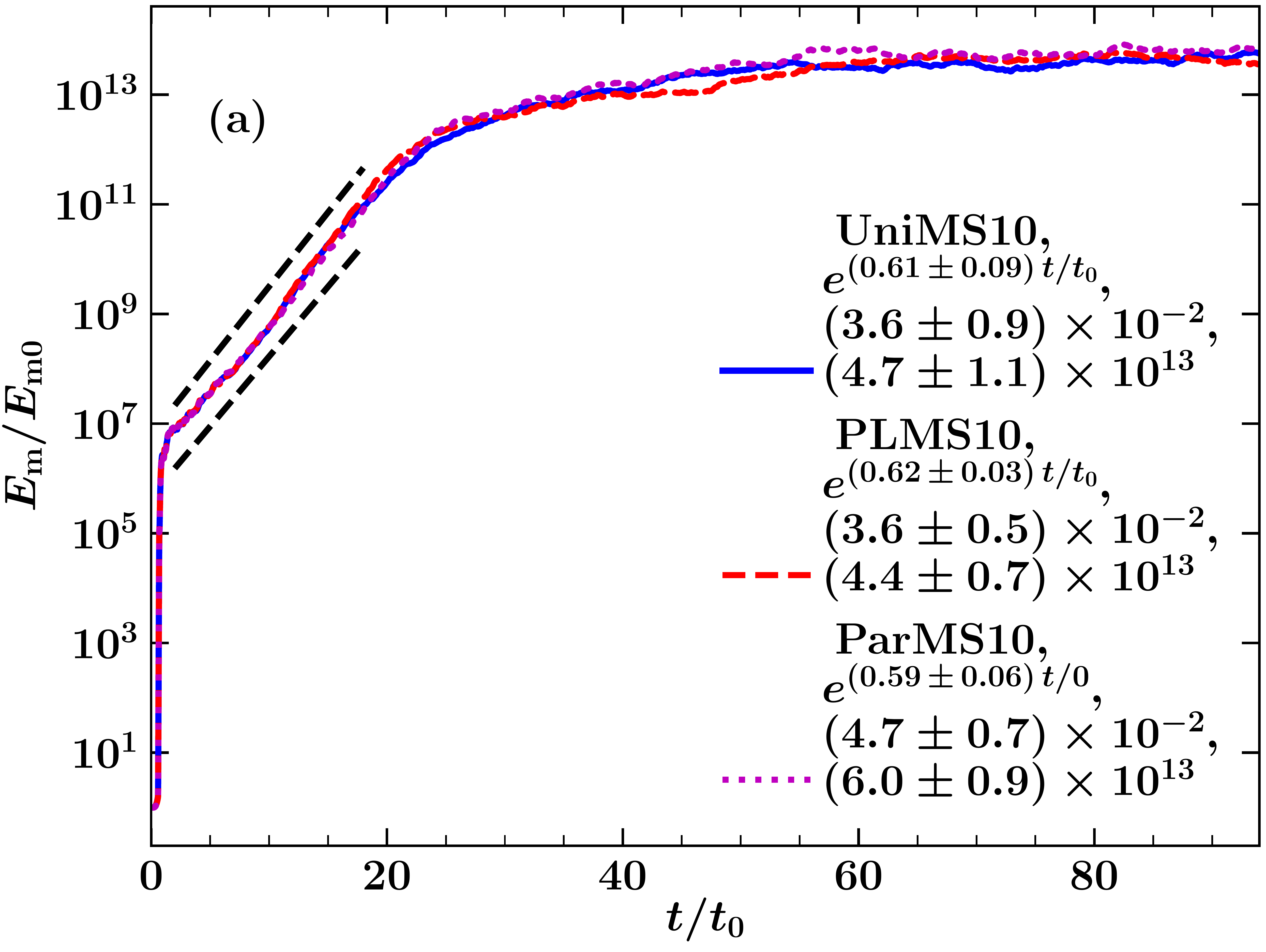}
\includegraphics[width=0.69\columnwidth]{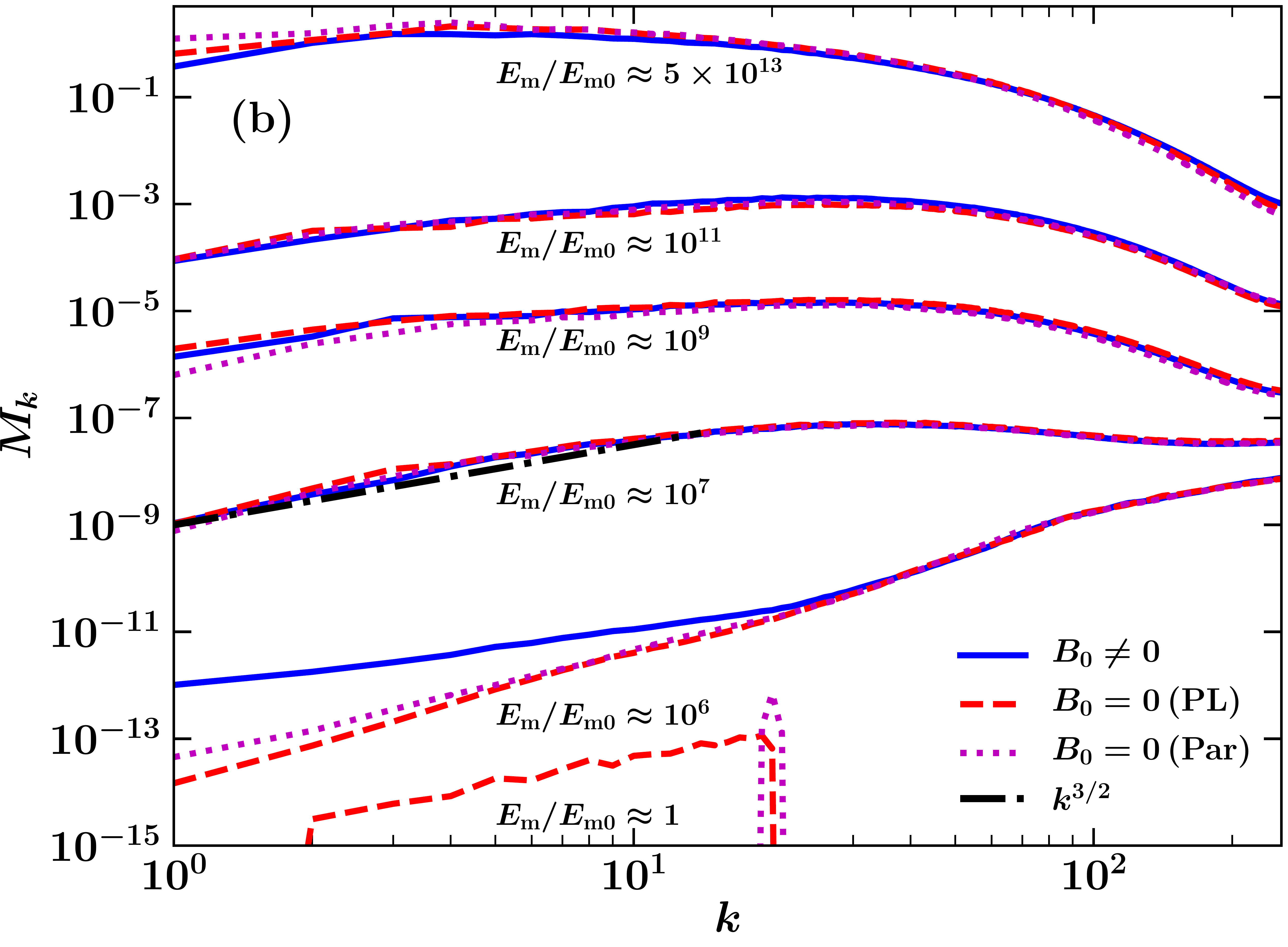}
\includegraphics[width=0.69\columnwidth]{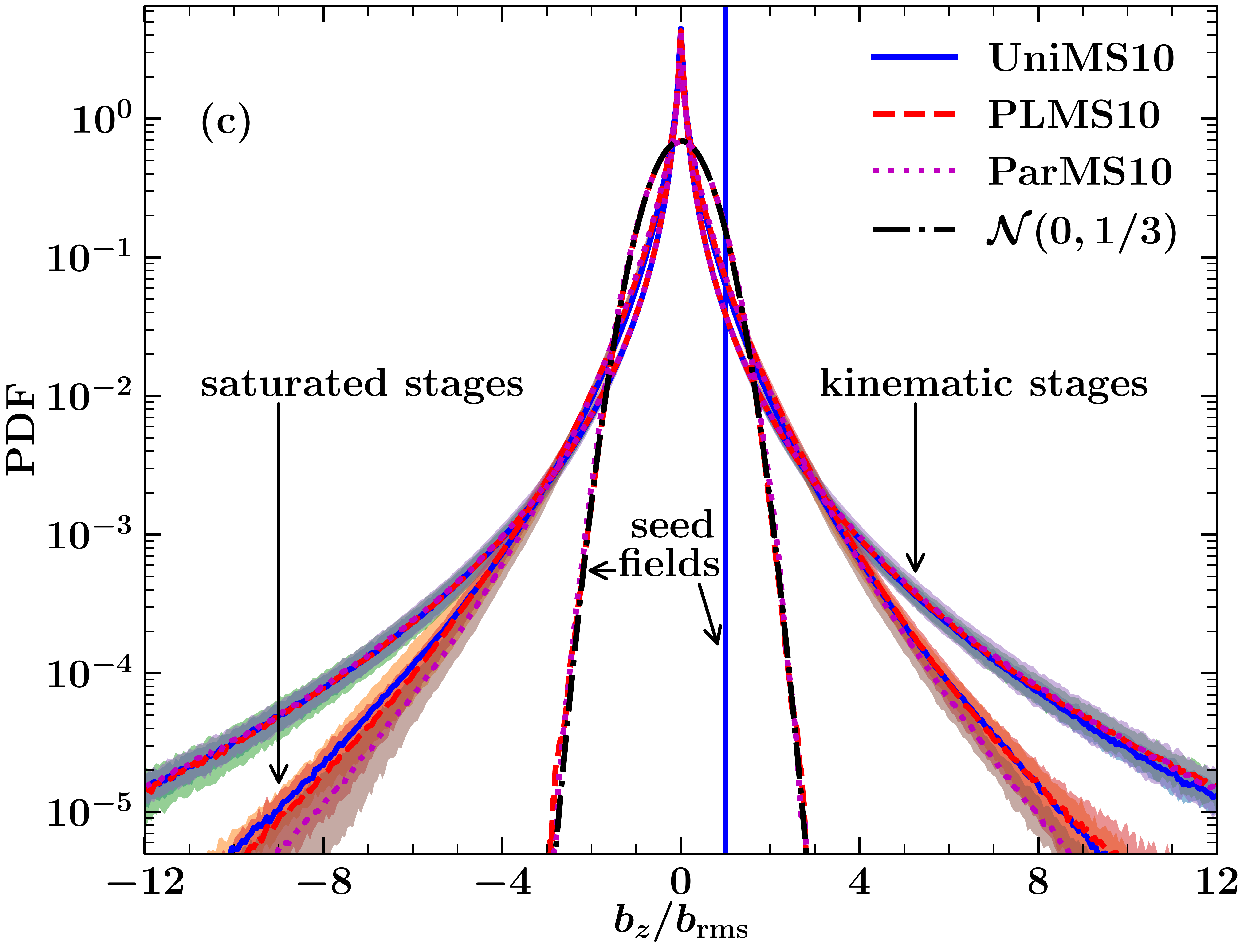}
\caption{Same as \Fig{b0}, but for $\Mach=10$ (models UniMS10, PLMS10, and ParMS10).}
\label{Mach}
\end{figure*}

In the previous subsections, we compared three different magnetic seed-field configurations, and showed that the resulting dynamo growth rates, saturation levels, and field properties are independent of the choice of the seed-field configuration. However, we only showed this for one particular set of MHD turbulence parameters, i.e., for an seed $\MachA=3.5\times10^5$, $\Mach=0.1$, and $\Re=\Rm=2000$, i.e., $\Pm=1$. Here we determine whether the same holds true when we vary these parameters for the dynamo in various additional simulations listed in \Tab{param}.

We show the results for varying the seed magnetic field strength or seed Alfv\'en Mach number in \Fig{b0} (models UniMA35000, PLMA35000, and ParMA35000), varying the magnetic Reynolds number in \Fig{rm} (models UniRm3000, PLRm3000, and ParRm3000), and varying the sonic Mach number in \Fig{Mach} (models UniMS10, PLMS10, and ParMS10). For the change in
the seed Alfv\'en Mach number and magnetic Reynolds number, the results (time evolution of magnetic energy, spectra, and PDFs of the $z$ component
of the magnetic field) follow similar trends as before.

The trends change slightly for $\Mach=10$. \Fig{Mach}(a) shows a very quick initial amplification of the magnetic field for all three seed-field cases, due to the compression of the magnetic field lines (not seen before for any $\Mach=0.1$ runs). This is also reflected in the spectra (\Fig{Mach}(b) for $E_{\rm m}/E_{\rm m0} \simeq 10^{6}$), where the fields on smaller scales are amplified more than on the larger scales (the power is non-zero on all scales for all three seed-field cases).  Once the magnetic field reaches the kinematic stage, the spectra for all three seed fields are equal and remain the same also in the saturated stage. The PDFs in the saturated stage are more intermittent than in the kinematic stage, but the PDFs both in the kinematic and saturated stages are more intermittent for the $\Mach=10$ case than for any of the $\Mach=0.1$ cases (compare the $x$-axis values and the heavy tails at higher values in \Fig{pdfs}a and \Fig{Mach}c). However, the growth rate and the structure of the magnetic fields in the kinematic and saturated stages, even for the $\Mach=10$ case, are unaffected by the configuration of the seed field.

The overall conclusions, even with changes in the plasma parameters, remain the same as before: the structure and strength of the seed magnetic fields do not affect the growth rate, final saturation level, shape of the magnetic power spectrum, and the statistical structure of the small-scale dynamo generated magnetic fields. Thus, the detailed properties of the seed field or even their various generation mechanisms are relatively unimportant when studying evolved magnetic fields that were amplified by the turbulent small-scale dynamo.

\section{Discussion} \label{dis}
\begin{figure}
\includegraphics[width=1.0\columnwidth]{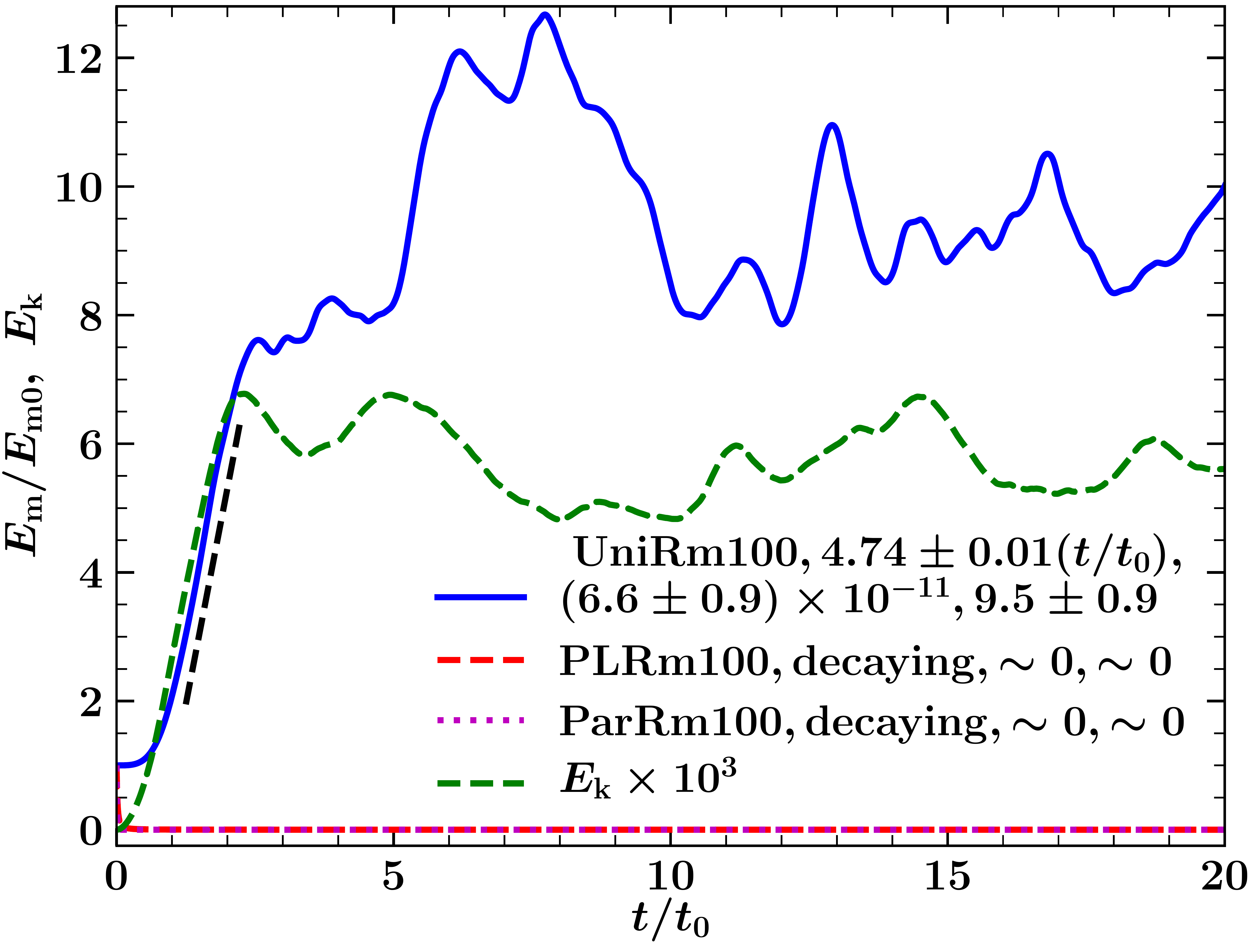}
\caption{Same as Fig.~\ref{ts}, but for models with a low magnetic Reynolds number ($\Rm=100$): UniRm100 (solid, blue), PLRm100 (dashed, red) and ParRm100 (dotted, magenta) and the time evolution of (scaled) turbulent kinetic energy $E_{\rm k}$ is also shown (dashed, green). Only the case with a uniform seed field shows initial amplification of the magnetic field, due to tangling (caused by the 2nd term on the right-hand side of Eq.~\ref{tmf}). The fitted line shows linear amplification (as opposed to dynamo amplification, which is exponential) with slope similar to that of the turbulent kinetic energy $E_{\rm k}$. None of these runs shows dynamo amplification, because the magnetic Reynolds number is less than the critical value.}
\label{rm100}
\end{figure}

From our results, we show that the seed magnetic field information is lost (within a few eddy turnover times, as seen in \Fig{ts}(a), \Fig{b0}(a), \Fig{rm}(a), and \Fig{Mach}(a)). Then the dynamo-generated magnetic field reaches a statistically similar state for all three seed fields, which is determined by the properties of the turbulence (such as $\Re$ and $\Mach$) and magnetic resistivity (or equivalently $\Rm$). This also shows that the small-scale dynamo is very efficient in amplifying (given that $\Rm$ is greater than its critical value, $\Rmc$) any non-zero seed field.

\subsection{Small-scale magnetic fields generated by the tangling of the large-scale field}

If $\Rm < \Rmc$, the small-scale dynamo is inactive. However, for the case with non-zero mean seed field (UniRm100), small-scale magnetic fields can still be generated by the tangling of the large-scale (or mean) uniform seed field \citep[see Appendix A in][for discussion on various mechanisms by which small-scale magnetic fields can be generated]{SetaEA2018}. Such a mechanism is only applicable in systems, such as spiral galaxies, where the large-scale field is maintained by other processes, i.e., $\alpha - \Omega$  large-scale dynamo \citep{KrauseR1980, ShukurovS2008, ChamandyEA2014} or cosmic-ray driven dynamo \citep{Parker1992, HanaszEA2004}. In our simulations, the large-scale uniform field is maintained at all times because the MHD equations (\Eq{ce}--\Eq{div}) preserve the net seed magnetic flux in a triply periodic numerical domain.
 
We demonstrate the growth of small-scale magnetic fields
by the tangling of the uniform seed field in \Fig{rm100} by choosing $\Rm=100 < \Rmc (\approx 165)$. In \Fig{rm100}(a), we show the evolution of magnetic and turbulent kinetic energies for models with all three seed field cases (UniRm100, PLRm100, and ParRm100). The magnetic energy only grows (linearly) for the case UniRm100 and decays for the other two cases (PLRm100 and ParRm100).

Analytically the total magnetic field ($\vec{B}$) in the induction equation can be divided into the small-scale, random ($\vec{b}$) and large-scale, mean ($\vec{B_0}$) field.
Substituting in \Eq{ie}, we get
\begin{align}
\frac{\partial (\vec{b} + \vec{B_0})}{\partial t} = \nabla \times (\vec{v} \times (\vec{b} + \vec{B_0})) + \eta \nabla^2 (\vec{b} + \vec{B_0}).  
\label{ietmf}
\end{align}
Since $\vec{B_0}$,  in comparison to $\vec{b}$,  varies over a much larger spatial and temporal scales (here, $\vec{B_0}$ is constant in space and time), we can simplify \Eq{ietmf} further as,
\begin{align}
\frac{\partial \vec{b}}{\partial t} = &\underbrace{\nabla \times (\vec{v} \times \vec{b})}_{\text {stretching and compression}} +  \underbrace{\nabla \times (\vec{v} \times \vec{B_0})}_{\text{tangling of the large-scale field}}  +  \nonumber \\ 
 &\underbrace{\eta \nabla^2 \vec{b}.}_{\text{magnetic diffusion}}
\label{tmf}
\end{align}
The first term on the right-hand side of the \Eq{tmf} represents the amplification
of the small-scale magnetic field due to the stretching and compression of magnetic field lines by the turbulent velocity. The second term on the right-hand side of the \Eq{tmf} represents amplification of the magnetic field due to the tangling of the large-scale field by the turbulent velocity. The last term in \Eq{tmf} represents magnetic diffusion. Initially, for a uniform seed field (UniRm100), the first and third terms in \Eq{tmf} are negligible since the small-scale field is very small (zero to start with) and the magnetic power is negligible on smaller scales where the diffusion is primarily active. Once the small-scale magnetic field grows, the other two terms in \Eq{tmf} can no longer be ignored, especially the diffusion term which becomes quite significant because the magnetic power is non-zero on smaller scales. The field eventually reaches a statistically steady-state caused by the competition between tangling and diffusion. 


\subsection{In the absence of active turbulence}

Our main conclusion from this study, namely that the strength and structure of the seed magnetic field do not affect dynamo amplification holds true only for regions with a significant level of turbulence. In contrast, for low-density regions with negligible turbulence, (e.g., voids of the large-scale cosmic structure), the evolved magnetic fields still have imprints of the seed field \citep[can be seen in the results from the IllustrisTNG cosmological simulations; see e.g., figures~~14 and~19 in][]{MarinacciEA2015}. From the observed limits of the magnetic field strengths in voids \citep{NeronovV2010, TavecchioEA2011, TiedeEA2020}, it is unclear whether they are primordial (early Universe) or developed later in the history of the Universe due to galactic winds \citep{BertoneVE2006,SamuiSS2018} or outflows from AGNs \citep{FurlanettoL2001}.

Observations of magnetic fields in young galaxies \citep{BernetEA2008,MaoEA2017,MalikCS2020} confirm the presence of strong $\muG$ fields, comparable to the magnetic field strengths observed in the Milky Way \citep{Haverkorn2015}, and in nearby spiral galaxies \citep{FletcherEA2004,FletcherEA2011,Beck2015,Beck2016}. The present observations of magnetic fields, even in young galaxies, cannot help us distinguish between various seed-field generation scenarios (especially astrophysical vs.~primordial) because any seed-field information is lost very early on in the evolution. However, primordial magnetic fields can still be studied via other observational probes \citep[see, e.g., table~1 in][]{Subramanian2016}.

\section{Conclusions} \label{con}
We studied the effect of the strength and structure of seed magnetic fields on the small-scale turbulent dynamo action via numerical simulations. Motivated by various seed-field generation mechanisms, we select three possible seed-field configurations: uniform field ($B_0 \ne 0$, Uni), random field with a power-law spectrum ($B_0 = 0$, PL), 
and random field with a parabolic spectrum ($B_0 = 0$, Par). Based on our results,
we arrive at the following conclusions:
\begin{itemize}
\item The magnetic field growth rate in the kinematic stage, and the final saturated amplification level are not affected by the strength and structure of the seed field (\Fig{ts}(a)).
\item The spectrum, though very different for all three seed-field cases, quickly spreads over a range of wave numbers, and is statistically similar in both the kinematic and saturated stages for all three cases (\Fig{spec}(a)). The computed coherence length of the magnetic field is lower in the kinematic stage as compared to the saturated stage, but is very similar for all three seed-field cases (\Fig{spec}(b)).
\item Even though all three seed fields have very different spatial structure (first row in \Fig{2dplots}), the magnetic field structure in the kinematic and saturated stages is statistically similar for all three seed-field cases (\Fig{2dplots}(b), \Fig{2dplots}(c), and \Fig{pdfs}).
\item Above conclusions are not altered by changing the seed field strength or the seed Alfv\'en Mach number (\Fig{b0}), magnetic Reynolds number $\Rm$ (\Fig{rm}), or turbulent sonic Mach number (\Fig{Mach}).
\item The seed-field information is lost, since the small-scale turbulent dynamo is very efficient in amplifying seed fields of any form, as long as it is non-zero and the magnetic Reynolds number is greater than its critical value. The details of the seed field or their various generation mechanisms are not important when studying the magnetic field amplified by the small-scale turbulent dynamo.
\item Even when the small-scale dynamo action is not active, a weak small-scale magnetic field can be generated and maintained by the tangling of the large-scale magnetic field (if present) (\Fig{rm100}). The tangled small-scale magnetic field in such a case only grows linearly and then saturates once the diffusion term balances the tangling term (Eq.~\ref{tmf}). 
\end{itemize}

\section*{Acknowledgements}
We thank the anonymous referee for a thorough review of this work. A.~S.~thanks Ramkishor Sharma for interesting discussions on primordial magnetic fields. C.~F.~acknowledges funding provided by the Australian Research Council (Discovery Project DP170100603 and Future Fellowship FT180100495), and the Australia-Germany Joint Research Cooperation Scheme (UA-DAAD). We further acknowledge high-performance computing resources provided by the Leibniz Rechenzentrum and the Gauss Centre for Supercomputing (grants~pr32lo, pr48pi and GCS Large-scale project~10391), the Australian National Computational Infrastructure (grant~ek9) in the framework of the National Computational Merit Allocation Scheme and the ANU Merit Allocation Scheme. The simulation software FLASH was in part developed by the DOE-supported Flash Center for Computational Science at the University of Chicago.

\section*{Data availability}
The data used in this article is available upon request to the corresponding author, Amit Seta (\href{mailto:amit.seta@anu.adu.au}{amit.seta@anu.adu.au}).



\bibliographystyle{mnras}
\bibliography{seedb} 

\begin{thebibliography}{}
\makeatletter
\relax
\def\mn@urlcharsother{\let\do\@makeother \do\$\do\&\do\#\do\^\do\_\do\%\do\~}
\def\mn@doi{\begingroup\mn@urlcharsother \@ifnextchar [ {\mn@doi@}
  {\mn@doi@[]}}
\def\mn@doi@[#1]#2{\def\@tempa{#1}\ifx\@tempa\@empty \href
  {http://dx.doi.org/#2} {doi:#2}\else \href {http://dx.doi.org/#2} {#1}\fi
  \endgroup}
\def\mn@eprint#1#2{\mn@eprint@#1:#2::\@nil}
\def\mn@eprint@arXiv#1{\href {http://arxiv.org/abs/#1} {{\tt arXiv:#1}}}
\def\mn@eprint@dblp#1{\href {http://dblp.uni-trier.de/rec/bibtex/#1.xml}
  {dblp:#1}}
\def\mn@eprint@#1:#2:#3:#4\@nil{\def\@tempa {#1}\def\@tempb {#2}\def\@tempc
  {#3}\ifx \@tempc \@empty \let \@tempc \@tempb \let \@tempb \@tempa \fi \ifx
  \@tempb \@empty \def\@tempb {arXiv}\fi \@ifundefined
  {mn@eprint@\@tempb}{\@tempb:\@tempc}{\expandafter \expandafter \csname
  mn@eprint@\@tempb\endcsname \expandafter{\@tempc}}}

\bibitem[\protect\citeauthoryear{{Batchelor}}{{Batchelor}}{1950}]{Batchelor1950}
{Batchelor} G.~K.,  1950, \mn@doi [Proceedings of the Royal Society of London
  Series A] {10.1098/rspa.1950.0069}, \href
  {https://ui.adsabs.harvard.edu/abs/1950RSPSA.201..405B} {201, 405}

\bibitem[\protect\citeauthoryear{{Beck}}{{Beck}}{2015}]{Beck2015}
{Beck} R.,  2015, \mn@doi [\aap] {10.1051/0004-6361/201425572}, \href
  {https://ui.adsabs.harvard.edu/abs/2015A&A...578A..93B} {578, A93}

\bibitem[\protect\citeauthoryear{{Beck}}{{Beck}}{2016}]{Beck2016}
{Beck} R.,  2016, \araa, \href
  {http://adsabs.harvard.edu/abs/2016A%26ARv..24....4B} {24, 4}

\bibitem[\protect\citeauthoryear{{Beck}, {Brandenburg}, {Moss}, {Shukurov}  \&
  {Sokoloff}}{{Beck} et~al.}{1996}]{BeckEA1996}
{Beck} R.,  {Brandenburg} A.,  {Moss} D.,  {Shukurov} A.,   {Sokoloff} D.,
  1996, \mn@doi [\araa] {10.1146/annurev.astro.34.1.155}, \href
  {http://adsabs.harvard.edu/abs/1996ARA%26A..34..155B} {34, 155}

\bibitem[\protect\citeauthoryear{{Beresnyak}}{{Beresnyak}}{2012}]{Beresnyak2012}
{Beresnyak} A.,  2012, \mn@doi [\prl] {10.1103/PhysRevLett.108.035002}, \href
  {https://ui.adsabs.harvard.edu/#abs/2012PhRvL.108c5002B} {108, 035002}

\bibitem[\protect\citeauthoryear{{Bernet}, {Miniati}, {Lilly}, {Kronberg}  \&
  {Dessauges-Zavadsky}}{{Bernet} et~al.}{2008}]{BernetEA2008}
{Bernet} M.~L.,  {Miniati} F.,  {Lilly} S.~J.,  {Kronberg} P.~P.,
  {Dessauges-Zavadsky} M.,  2008, \mn@doi [\nat] {10.1038/nature07105}, \href
  {https://ui.adsabs.harvard.edu/#abs/2008Natur.454..302B} {454, 302}

\bibitem[\protect\citeauthoryear{{Bertone}, {Vogt}  \& {En{\ss}lin}}{{Bertone}
  et~al.}{2006}]{BertoneVE2006}
{Bertone} S.,  {Vogt} C.,   {En{\ss}lin} T.,  2006, \mn@doi [\mnras]
  {10.1111/j.1365-2966.2006.10474.x}, \href
  {https://ui.adsabs.harvard.edu/abs/2006MNRAS.370..319B} {370, 319}

\bibitem[\protect\citeauthoryear{{Bhat} \& {Subramanian}}{{Bhat} \&
  {Subramanian}}{2013}]{BhatS2013}
{Bhat} P.,  {Subramanian} K.,  2013, \mn@doi [\mnras] {10.1093/mnras/sts516},
  \href {https://ui.adsabs.harvard.edu/#abs/2013MNRAS.429.2469B} {429, 2469}

\bibitem[\protect\citeauthoryear{{Bhat} \& {Subramanian}}{{Bhat} \&
  {Subramanian}}{2014}]{BhatS2014}
{Bhat} P.,  {Subramanian} K.,  2014, \mn@doi [\apj]
  {10.1088/2041-8205/791/2/L34}, \href
  {https://ui.adsabs.harvard.edu/#abs/2014ApJ...791L..34B} {791, L34}

\bibitem[\protect\citeauthoryear{{Biermann}}{{Biermann}}{1950}]{Biermann1950}
{Biermann} L.,  1950, Zeitschrift Naturforschung Teil A, \href
  {https://ui.adsabs.harvard.edu/abs/1950ZNatA...5...65B} {5, 65}

\bibitem[\protect\citeauthoryear{{Boldyrev} \& {Cattaneo}}{{Boldyrev} \&
  {Cattaneo}}{2004}]{BoldyrevC2004}
{Boldyrev} S.,  {Cattaneo} F.,  2004, \mn@doi [\prl]
  {10.1103/PhysRevLett.92.144501}, \href
  {https://ui.adsabs.harvard.edu/#abs/2004PhRvL..92n4501B} {92, 144501}

\bibitem[\protect\citeauthoryear{Bouchut, Klingenberg  \& Waagan}{Bouchut
  et~al.}{2007}]{BouchutKW2007}
Bouchut F.,  Klingenberg C.,   Waagan K.,  2007, Numerische Mathematik, 108, 7

\bibitem[\protect\citeauthoryear{Bouchut, Klingenberg  \& Waagan}{Bouchut
  et~al.}{2010}]{BouchutKW2010}
Bouchut F.,  Klingenberg C.,   Waagan K.,  2010, Numerische Mathematik, 115,
  647

\bibitem[\protect\citeauthoryear{{Brandenburg} \& {Subramanian}}{{Brandenburg}
  \& {Subramanian}}{2005}]{BrandenburgS2005}
{Brandenburg} A.,  {Subramanian} K.,  2005, \mn@doi [\physrep]
  {10.1016/j.physrep.2005.06.005}, \href
  {http://adsabs.harvard.edu/abs/2005PhR...417....1B} {417, 1}

\bibitem[\protect\citeauthoryear{{Brandenburg}, {Kahniashvili}  \&
  {Tevzadze}}{{Brandenburg} et~al.}{2015}]{BrandenburgKT2015}
{Brandenburg} A.,  {Kahniashvili} T.,   {Tevzadze} A. e.~G.,  2015, \mn@doi
  [\prl] {10.1103/PhysRevLett.114.075001}, \href
  {https://ui.adsabs.harvard.edu/abs/2015PhRvL.114g5001B} {114, 075001}

\bibitem[\protect\citeauthoryear{{Brandenburg}, {Haugen}, {Li}  \&
  {Subramanian}}{{Brandenburg} et~al.}{2018}]{BrandenburgEA2018}
{Brandenburg} A.,  {Haugen} N.~E.~L.,  {Li} X.-Y.,   {Subramanian} K.,  2018,
  \mn@doi [\mnras] {10.1093/mnras/sty1570}, \href
  {https://ui.adsabs.harvard.edu/#abs/2018MNRAS.479.2827B} {479, 2827}

\bibitem[\protect\citeauthoryear{{Bushby} \& {Favier}}{{Bushby} \&
  {Favier}}{2014}]{BushbyF2014}
{Bushby} P.~J.,  {Favier} B.,  2014, \mn@doi [\aap]
  {10.1051/0004-6361/201322993}, \href
  {https://ui.adsabs.harvard.edu/abs/2014A&A...562A..72B} {562, A72}

\bibitem[\protect\citeauthoryear{{Bushby}, {Proctor}  \& {Weiss}}{{Bushby}
  et~al.}{2010}]{BushbyPW2010}
{Bushby} P.~J.,  {Proctor} M.~R.~E.,   {Weiss} N.~O.,  2010, in Numerical
  Modeling of Space Plasma Flows, Astronum-2009. p.~181

\bibitem[\protect\citeauthoryear{{Carilli} \& {Taylor}}{{Carilli} \&
  {Taylor}}{2002}]{CarilliT2002}
{Carilli} C.~L.,  {Taylor} G.~B.,  2002, \mn@doi [\araa]
  {10.1146/annurev.astro.40.060401.093852}, \href
  {https://ui.adsabs.harvard.edu/abs/2002ARA&A..40..319C} {40, 319}

\bibitem[\protect\citeauthoryear{{Cattaneo}}{{Cattaneo}}{1999}]{Cattaneo1999}
{Cattaneo} F.,  1999, \mn@doi [\apj] {10.1086/311962}, \href
  {https://ui.adsabs.harvard.edu/#abs/1999ApJ...515L..39C} {515, L39}

\bibitem[\protect\citeauthoryear{{Cattaneo} \& {Tobias}}{{Cattaneo} \&
  {Tobias}}{2009}]{CattaneoT2009}
{Cattaneo} F.,  {Tobias} S.~M.,  2009, \mn@doi [Journal of Fluid Mechanics]
  {10.1017/S0022112008004990}, \href
  {https://ui.adsabs.harvard.edu/#abs/2009JFM...621..205C} {621, 205}

\bibitem[\protect\citeauthoryear{{Chamandy}, {Shukurov}, {Subramanian}  \&
  {Stoker}}{{Chamandy} et~al.}{2014}]{ChamandyEA2014}
{Chamandy} L.,  {Shukurov} A.,  {Subramanian} K.,   {Stoker} K.,  2014, \mn@doi
  [\mnras] {10.1093/mnras/stu1274}, \href
  {http://adsabs.harvard.edu/abs/2014MNRAS.443.1867C} {443, 1867}

\bibitem[\protect\citeauthoryear{{Cho}}{{Cho}}{2014}]{Cho2014}
{Cho} J.,  2014, \mn@doi [\apj] {10.1088/0004-637X/797/2/133}, \href
  {https://ui.adsabs.harvard.edu/abs/2014ApJ...797..133C} {797, 133}

\bibitem[\protect\citeauthoryear{{Cho} \& {Ryu}}{{Cho} \&
  {Ryu}}{2009}]{ChoR2009}
{Cho} J.,  {Ryu} D.,  2009, \mn@doi [\apj] {10.1088/0004-637X/705/1/L90}, \href
  {https://ui.adsabs.harvard.edu/#abs/2009ApJ...705L..90C} {705, L90}

\bibitem[\protect\citeauthoryear{{Cho}, {Vishniac}, {Beresnyak}, {Lazarian}  \&
  {Ryu}}{{Cho} et~al.}{2009}]{ChoEA2009}
{Cho} J.,  {Vishniac} E.~T.,  {Beresnyak} A.,  {Lazarian} A.,   {Ryu} D.,
  2009, \mn@doi [\apj] {10.1088/0004-637X/693/2/1449}, \href
  {https://ui.adsabs.harvard.edu/abs/2009ApJ...693.1449C} {693, 1449}

\bibitem[\protect\citeauthoryear{{Davies} \& {Widrow}}{{Davies} \&
  {Widrow}}{2000}]{DaviesW2000}
{Davies} G.,  {Widrow} L.~M.,  2000, \mn@doi [\apj] {10.1086/309358}, \href
  {https://ui.adsabs.harvard.edu/abs/2000ApJ...540..755D} {540, 755}

\bibitem[\protect\citeauthoryear{{Dom{\'i}nguez-Fern{\'a}ndez}, {Vazza},
  {Br{\"u}ggen}  \& {Brunetti}}{{Dom{\'i}nguez-Fern{\'a}ndez}
  et~al.}{2019}]{DominguezEA2019}
{Dom{\'i}nguez-Fern{\'a}ndez} P.,  {Vazza} F.,  {Br{\"u}ggen} M.,   {Brunetti}
  G.,  2019, \mn@doi [\mnras] {10.1093/mnras/stz877}, \href
  {https://ui.adsabs.harvard.edu/abs/2019MNRAS.486..623D} {486, 623}

\bibitem[\protect\citeauthoryear{{Dubey} et~al.,}{{Dubey}
  et~al.}{2008}]{DubeyEA2008}
{Dubey} A.,  et~al., 2008, {Challenges of Extreme Computing using the FLASH
  code}.
p.~145

\bibitem[\protect\citeauthoryear{{Durrer} \& {Neronov}}{{Durrer} \&
  {Neronov}}{2013}]{DurrerN2013}
{Durrer} R.,  {Neronov} A.,  2013, \mn@doi [\aapr] {10.1007/s00159-013-0062-7},
  \href {https://ui.adsabs.harvard.edu/abs/2013A&ARv..21...62D} {21, 62}

\bibitem[\protect\citeauthoryear{{Elmegreen}}{{Elmegreen}}{2009}]{Elmegreen2009}
{Elmegreen} B.~G.,  2009, in {Andersen} J.,  {Nordstr{\"o}ara} {m} B.,   {Bland
  -Hawthorn} J.,  eds,  IAU Symposium Vol. 254, The Galaxy Disk in Cosmological
  Context. pp 289--300 (\mn@eprint {arXiv} {0810.5406}),
  \mn@doi{10.1017/S1743921308027713}

\bibitem[\protect\citeauthoryear{{Eswaran} \& {Pope}}{{Eswaran} \&
  {Pope}}{1988}]{EswaranPope1988}
{Eswaran} V.,  {Pope} S.~B.,  1988, Computers and Fluids, \href
  {https://ui.adsabs.harvard.edu/abs/1988CF.....16..257E} {16, 257}

\bibitem[\protect\citeauthoryear{{Favier} \& {Bushby}}{{Favier} \&
  {Bushby}}{2012}]{FavierB2012}
{Favier} B.,  {Bushby} P.~J.,  2012, \mn@doi [Journal of Fluid Mechanics]
  {10.1017/jfm.2011.429}, \href
  {https://ui.adsabs.harvard.edu/#abs/2012JFM...690..262F} {690, 262}

\bibitem[\protect\citeauthoryear{{Federrath}}{{Federrath}}{2013}]{Federrath2013}
{Federrath} C.,  2013, \mn@doi [\mnras] {10.1093/mnras/stt1644}, \href
  {https://ui.adsabs.harvard.edu/#abs/2013MNRAS.436.1245F} {436, 1245}

\bibitem[\protect\citeauthoryear{{Federrath}}{{Federrath}}{2016}]{Federrath2016}
{Federrath} C.,  2016, \mn@doi [Journal of Plasma Physics]
  {10.1017/S0022377816001069}, \href
  {https://ui.adsabs.harvard.edu/#abs/2016JPlPh..82f5301F} {82, 535820601}

\bibitem[\protect\citeauthoryear{{Federrath}, {Roman-Duval}, {Klessen},
  {Schmidt}  \& {Mac Low}}{{Federrath} et~al.}{2010}]{FederrathEA2010}
{Federrath} C.,  {Roman-Duval} J.,  {Klessen} R.~S.,  {Schmidt} W.,   {Mac Low}
  M.-M.,  2010, \mn@doi [\aap] {10.1051/0004-6361/200912437}, \href
  {http://adsabs.harvard.edu/abs/2010A%26A...512A..81F} {512, A81}

\bibitem[\protect\citeauthoryear{{Federrath}, {Chabrier}, {Schober},
  {Banerjee}, {Klessen}  \& {Schleicher}}{{Federrath}
  et~al.}{2011}]{FederrathEA2011}
{Federrath} C.,  {Chabrier} G.,  {Schober} J.,  {Banerjee} R.,  {Klessen}
  R.~S.,   {Schleicher} D.~R.~G.,  2011, \mn@doi [\prl]
  {10.1103/PhysRevLett.107.114504}, \href
  {https://ui.adsabs.harvard.edu/#abs/2011PhRvL.107k4504F} {107, 114504}

\bibitem[\protect\citeauthoryear{{Federrath}, {Schober}, {Bovino}  \&
  {Schleicher}}{{Federrath} et~al.}{2014}]{FederrathEA2014}
{Federrath} C.,  {Schober} J.,  {Bovino} S.,   {Schleicher} D. R.~G.,  2014,
  \mn@doi [\apj] {10.1088/2041-8205/797/2/L19}, \href
  {https://ui.adsabs.harvard.edu/#abs/2014ApJ...797L..19F} {797, L19}

\bibitem[\protect\citeauthoryear{{Federrath} et~al.,}{{Federrath}
  et~al.}{2017}]{FederrathEA2017}
{Federrath} C.,  et~al., 2017, in {Crocker} R.~M.,  {Longmore} S.~N.,
  {Bicknell} G.~V.,  eds,  IAU Symposium Vol. 322, The Multi-Messenger
  Astrophysics of the Galactic Centre. pp 123--128 (\mn@eprint {arXiv}
  {1609.08726}), \mn@doi{10.1017/S1743921316012357}

\bibitem[\protect\citeauthoryear{{Fletcher}, {Berkhuijsen}, {Beck}  \&
  {Shukurov}}{{Fletcher} et~al.}{2004}]{FletcherEA2004}
{Fletcher} A.,  {Berkhuijsen} E.~M.,  {Beck} R.,   {Shukurov} A.,  2004,
  \mn@doi [\aap] {10.1051/0004-6361:20034133}, \href
  {https://ui.adsabs.harvard.edu/abs/2004A&A...414...53F} {414, 53}

\bibitem[\protect\citeauthoryear{{Fletcher}, {Beck}, {Shukurov}, {Berkhuijsen}
  \& {Horellou}}{{Fletcher} et~al.}{2011}]{FletcherEA2011}
{Fletcher} A.,  {Beck} R.,  {Shukurov} A.,  {Berkhuijsen} E.~M.,   {Horellou}
  C.,  2011, \mn@doi [\mnras] {10.1111/j.1365-2966.2010.18065.x}, \href
  {https://ui.adsabs.harvard.edu/#abs/2011MNRAS.412.2396F} {412, 2396}

\bibitem[\protect\citeauthoryear{{Fryxell} et~al.,}{{Fryxell}
  et~al.}{2000}]{FryxellEA2000}
{Fryxell} B.,  et~al., 2000, \mn@doi [\apjs] {10.1086/317361}, \href
  {https://ui.adsabs.harvard.edu/abs/2000ApJS..131..273F} {131, 273}

\bibitem[\protect\citeauthoryear{{Furlanetto} \& {Loeb}}{{Furlanetto} \&
  {Loeb}}{2001}]{FurlanettoL2001}
{Furlanetto} S.~R.,  {Loeb} A.,  2001, \mn@doi [\apj] {10.1086/321630}, \href
  {https://ui.adsabs.harvard.edu/abs/2001ApJ...556..619F} {556, 619}

\bibitem[\protect\citeauthoryear{{Gilbert}, {Mason}  \& {Tobias}}{{Gilbert}
  et~al.}{2016}]{GilbertMT2016}
{Gilbert} A.~D.,  {Mason} J.,   {Tobias} S.~M.,  2016, \mn@doi [Journal of
  Fluid Mechanics] {10.1017/jfm.2016.60}, \href
  {https://ui.adsabs.harvard.edu/#abs/2016JFM...791..568G} {791, 568}

\bibitem[\protect\citeauthoryear{{Gnedin}, {Ferrara}  \& {Zweibel}}{{Gnedin}
  et~al.}{2000}]{GnedinFZ2000}
{Gnedin} N.~Y.,  {Ferrara} A.,   {Zweibel} E.~G.,  2000, \mn@doi [\apj]
  {10.1086/309272}, \href
  {https://ui.adsabs.harvard.edu/abs/2000ApJ...539..505G} {539, 505}

\bibitem[\protect\citeauthoryear{{Govoni} \& {Feretti}}{{Govoni} \&
  {Feretti}}{2004}]{GovoniF2004}
{Govoni} F.,  {Feretti} L.,  2004, \mn@doi [International Journal of Modern
  Physics D] {10.1142/S0218271804005080}, \href
  {https://ui.adsabs.harvard.edu/#abs/2004IJMPD..13.1549G} {13, 1549}

\bibitem[\protect\citeauthoryear{{Hanasz}, {Kowal}, {Otmianowska-Mazur}  \&
  {Lesch}}{{Hanasz} et~al.}{2004}]{HanaszEA2004}
{Hanasz} M.,  {Kowal} G.,  {Otmianowska-Mazur} K.,   {Lesch} H.,  2004, \mn@doi
  [\apj] {10.1086/420697}, \href
  {https://ui.adsabs.harvard.edu/#abs/2004ApJ...605L..33H} {605, L33}

\bibitem[\protect\citeauthoryear{{Haugen}, {Brandenburg}  \& {Dobler}}{{Haugen}
  et~al.}{2004}]{HaugenBD2004}
{Haugen} N.~E.,  {Brandenburg} A.,   {Dobler} W.,  2004, \mn@doi [\pre]
  {10.1103/PhysRevE.70.016308}, \href
  {http://adsabs.harvard.edu/abs/2004PhRvE..70a6308H} {70, 016308}

\bibitem[\protect\citeauthoryear{{Haverkorn}}{{Haverkorn}}{2015}]{Haverkorn2015}
{Haverkorn} M.,  2015, in {Lazarian} A.,  {de Gouveia Dal Pino} E.~M.,
  {Melioli} C.,  eds,  Astrophysics and Space Science Library Vol. 407,
  Magnetic Fields in Diffuse Media. p.~483 (\mn@eprint {arXiv} {1406.0283}),
  \mn@doi{10.1007/978-3-662-44625-6_17}

\bibitem[\protect\citeauthoryear{{Haverkorn}, {Brown}, {Gaensler}  \&
  {McClure-Griffiths}}{{Haverkorn} et~al.}{2008}]{HaverkornEA2008}
{Haverkorn} M.,  {Brown} J.~C.,  {Gaensler} B.~M.,   {McClure-Griffiths} N.~M.,
   2008, \mn@doi [\apj] {10.1086/587165}, \href
  {http://adsabs.harvard.edu/abs/2008ApJ...680..362H} {680, 362}

\bibitem[\protect\citeauthoryear{{Iskakov}, {Schekochihin}, {Cowley},
  {McWilliams}  \& {Proctor}}{{Iskakov} et~al.}{2007}]{IskakovEA2007}
{Iskakov} A.~B.,  {Schekochihin} A.~A.,  {Cowley} S.~C.,  {McWilliams} J.~C.,
  {Proctor} M.~R.~E.,  2007, \mn@doi [\prl] {10.1103/PhysRevLett.98.208501},
  \href {https://ui.adsabs.harvard.edu/#abs/2007PhRvL..98t8501I} {98, 208501}

\bibitem[\protect\citeauthoryear{{Kazantsev}}{{Kazantsev}}{1968}]{Kazantsev1968}
{Kazantsev} A.~P.,  1968, Soviet Journal of Experimental and Theoretical
  Physics, \href {http://adsabs.harvard.edu/abs/1968JETP...26.1031K} {26, 1031}

\bibitem[\protect\citeauthoryear{{Krause} \& {R\"{a}dler}}{{Krause} \&
  {R\"{a}dler}}{1980}]{KrauseR1980}
{Krause} F.,  {R\"{a}dler} K.~H.,  1980, {Mean-field magnetohydrodynamics and
  dynamo theory}

\bibitem[\protect\citeauthoryear{{Krumholz} \& {Burkhart}}{{Krumholz} \&
  {Burkhart}}{2016}]{KrumholzB2016}
{Krumholz} M.~R.,  {Burkhart} B.,  2016, \mn@doi [\mnras]
  {10.1093/mnras/stw434}, \href
  {https://ui.adsabs.harvard.edu/abs/2016MNRAS.458.1671K} {458, 1671}

\bibitem[\protect\citeauthoryear{{Krumholz}, {Burkhart}, {Forbes}  \&
  {Crocker}}{{Krumholz} et~al.}{2018}]{KrumholzEA2018}
{Krumholz} M.~R.,  {Burkhart} B.,  {Forbes} J.~C.,   {Crocker} R.~M.,  2018,
  \mn@doi [\mnras] {10.1093/mnras/sty852}, \href
  {https://ui.adsabs.harvard.edu/abs/2018MNRAS.477.2716K} {477, 2716}

\bibitem[\protect\citeauthoryear{{Kulsrud} \& {Anderson}}{{Kulsrud} \&
  {Anderson}}{1992}]{KulsrudA1992}
{Kulsrud} R.~M.,  {Anderson} S.~W.,  1992, \mn@doi [\apj] {10.1086/171743},
  \href {http://adsabs.harvard.edu/abs/1992ApJ...396..606K} {396, 606}

\bibitem[\protect\citeauthoryear{{Kulsrud} \& {Zweibel}}{{Kulsrud} \&
  {Zweibel}}{2008}]{KulsrudZ08}
{Kulsrud} R.~M.,  {Zweibel} E.~G.,  2008, \mn@doi [Reports on Progress in
  Physics] {10.1088/0034-4885/71/4/046901}, \href
  {https://ui.adsabs.harvard.edu/#abs/2008RPPh...71d6901K} {71}

\bibitem[\protect\citeauthoryear{{Kulsrud}, {Cowley}, {Gruzinov}  \&
  {Sudan}}{{Kulsrud} et~al.}{1997a}]{KulsrudEA1997a}
{Kulsrud} R.,  {Cowley} S.~C.,  {Gruzinov} A.~V.,   {Sudan} R.~N.,  1997a,
  \mn@doi [\physrep] {10.1016/S0370-1573(96)00061-0}, \href
  {https://ui.adsabs.harvard.edu/abs/1997PhR...283..213K} {283, 213}

\bibitem[\protect\citeauthoryear{{Kulsrud}, {Cen}, {Ostriker}  \&
  {Ryu}}{{Kulsrud} et~al.}{1997b}]{KulsrudEA1997b}
{Kulsrud} R.~M.,  {Cen} R.,  {Ostriker} J.~P.,   {Ryu} D.,  1997b, \mn@doi
  [\apj] {10.1086/303987}, \href
  {https://ui.adsabs.harvard.edu/abs/1997ApJ...480..481K} {480, 481}

\bibitem[\protect\citeauthoryear{{Langer} \& {Durrive}}{{Langer} \&
  {Durrive}}{2018}]{LangerM2018}
{Langer} M.,  {Durrive} J.-B.,  2018, \mn@doi [Galaxies]
  {10.3390/galaxies6040124}, \href
  {https://ui.adsabs.harvard.edu/abs/2018Galax...6..124L} {6, 124}

\bibitem[\protect\citeauthoryear{{Mac Low} \& {Klessen}}{{Mac Low} \&
  {Klessen}}{2004}]{MacLowK2004}
{Mac Low} M.-M.,  {Klessen} R.~S.,  2004, \mn@doi [Reviews of Modern Physics]
  {10.1103/RevModPhys.76.125}, \href
  {https://ui.adsabs.harvard.edu/abs/2004RvMP...76..125M} {76, 125}

\bibitem[\protect\citeauthoryear{{Malik}, {Chand}  \& {Seshadri}}{{Malik}
  et~al.}{2020}]{MalikCS2020}
{Malik} S.,  {Chand} H.,   {Seshadri} T.~R.,  2020, \mn@doi [\apj]
  {10.3847/1538-4357/ab6bd5}, \href
  {https://ui.adsabs.harvard.edu/abs/2020ApJ...890..132M} {890, 132}

\bibitem[\protect\citeauthoryear{{Mao} et~al.,}{{Mao} et~al.}{2017}]{MaoEA2017}
{Mao} S.~A.,  et~al., 2017, \mn@doi [Nature Astronomy]
  {10.1038/s41550-017-0218-x}, \href
  {https://ui.adsabs.harvard.edu/#abs/2017NatAs...1..621M} {1, 621}

\bibitem[\protect\citeauthoryear{{Marinacci}, {Vogelsberger}, {Mocz}  \&
  {Pakmor}}{{Marinacci} et~al.}{2015}]{MarinacciEA2015}
{Marinacci} F.,  {Vogelsberger} M.,  {Mocz} P.,   {Pakmor} R.,  2015, \mn@doi
  [\mnras] {10.1093/mnras/stv1692}, \href
  {https://ui.adsabs.harvard.edu/abs/2015MNRAS.453.3999M} {453, 3999}

\bibitem[\protect\citeauthoryear{{Marinacci} et~al.,}{{Marinacci}
  et~al.}{2018}]{MarinacciEA2018}
{Marinacci} F.,  et~al., 2018, \mn@doi [\mnras] {10.1093/mnras/sty2206}, \href
  {https://ui.adsabs.harvard.edu/abs/2018MNRAS.480.5113M} {480, 5113}

\bibitem[\protect\citeauthoryear{Martins~Afonso, Mitra  \&
  Vincenzi}{Martins~Afonso et~al.}{2019}]{AfonsoMV2019}
Martins~Afonso M.,  Mitra D.,   Vincenzi D.,  2019, \mn@doi [Proceedings of the
  Royal Society A: Mathematical, Physical and Engineering Sciences]
  {10.1098/rspa.2018.0591}, 475, 20180591

\bibitem[\protect\citeauthoryear{{Meneguzzi}, {Frisch}  \&
  {Pouquet}}{{Meneguzzi} et~al.}{1981}]{MeneguzziFP1981}
{Meneguzzi} M.,  {Frisch} U.,   {Pouquet} A.,  1981, \mn@doi [Physical Review
  Letters] {10.1103/PhysRevLett.47.1060}, \href
  {http://adsabs.harvard.edu/abs/1981PhRvL..47.1060M} {47, 1060}

\bibitem[\protect\citeauthoryear{{Moffatt}}{{Moffatt}}{1978}]{Moffatt1978}
{Moffatt} H.~K.,  1978, {Magnetic field generation in electrically conducting
  fluids.}.
Cambridge University Press, Cambridge

\bibitem[\protect\citeauthoryear{{Neronov} \& {Vovk}}{{Neronov} \&
  {Vovk}}{2010}]{NeronovV2010}
{Neronov} A.,  {Vovk} I.,  2010, \mn@doi [Science] {10.1126/science.1184192},
  \href {https://ui.adsabs.harvard.edu/abs/2010Sci...328...73N} {328, 73}

\bibitem[\protect\citeauthoryear{{Pakmor} et~al.,}{{Pakmor}
  et~al.}{2017}]{PakmorEA2017}
{Pakmor} R.,  et~al., 2017, \mn@doi [\mnras] {10.1093/mnras/stx1074}, \href
  {https://ui.adsabs.harvard.edu/#abs/2017MNRAS.469.3185P} {469, 3185}

\bibitem[\protect\citeauthoryear{{Parker}}{{Parker}}{1992}]{Parker1992}
{Parker} E.~N.,  1992, \mn@doi [\apj] {10.1086/172046}, \href
  {https://ui.adsabs.harvard.edu/#abs/1992ApJ...401..137P} {401, 137}

\bibitem[\protect\citeauthoryear{{Rees}}{{Rees}}{2005}]{Rees2005}
{Rees} M.~J.,  2005, {Magnetic Fields in the Early Universe}.
p.~1, \mn@doi{10.1007/11369875_1}

\bibitem[\protect\citeauthoryear{{Rieder} \& {Teyssier}}{{Rieder} \&
  {Teyssier}}{2016}]{RiederT2016}
{Rieder} M.,  {Teyssier} R.,  2016, \mn@doi [\mnras] {10.1093/mnras/stv2985},
  \href {https://ui.adsabs.harvard.edu/abs/2016MNRAS.457.1722R} {457, 1722}

\bibitem[\protect\citeauthoryear{{Rieder} \& {Teyssier}}{{Rieder} \&
  {Teyssier}}{2017a}]{RiederT2017}
{Rieder} M.,  {Teyssier} R.,  2017a, \mn@doi [\mnras] {10.1093/mnras/stx1670},
  \href {https://ui.adsabs.harvard.edu/abs/2017MNRAS.471.2674R} {471, 2674}

\bibitem[\protect\citeauthoryear{{Rieder} \& {Teyssier}}{{Rieder} \&
  {Teyssier}}{2017b}]{RiederT2017b}
{Rieder} M.,  {Teyssier} R.,  2017b, \mn@doi [\mnras] {10.1093/mnras/stx2276},
  \href {https://ui.adsabs.harvard.edu/abs/2017MNRAS.472.4368R} {472, 4368}

\bibitem[\protect\citeauthoryear{{Rincon}}{{Rincon}}{2019}]{Rincon2019}
{Rincon} F.,  2019, \mn@doi [Journal of Plasma Physics]
  {10.1017/S0022377819000539}, \href
  {https://ui.adsabs.harvard.edu/abs/2019JPlPh..85d2001R} {85, 205850401}

\bibitem[\protect\citeauthoryear{{Ruzmaikin}, {Sokoloff}  \&
  {Shukurov}}{{Ruzmaikin} et~al.}{1988}]{RuzmaikinSS1988}
{Ruzmaikin} A.~A.,  {Sokoloff} D.~D.,   {Shukurov} A.~M.,  eds, 1988, {Magnetic
  fields of galaxies}  Astrophysics and Space Science Library Vol. 133,
  \mn@doi{10.1007/978-94-009-2835-0.
}

\bibitem[\protect\citeauthoryear{{Ruzmaikin}, {Sokoloff}  \&
  {Shukurov}}{{Ruzmaikin} et~al.}{1989}]{RuzmaikinSS1989}
{Ruzmaikin} A.,  {Sokoloff} D.,   {Shukurov} A.,  1989, \mn@doi [\mnras]
  {10.1093/mnras/241.1.1}, \href
  {http://adsabs.harvard.edu/abs/1989MNRAS.241....1R} {241, 1}

\bibitem[\protect\citeauthoryear{{Samui}, {Subramanian}  \& {Srianand
  }}{{Samui} et~al.}{2018}]{SamuiSS2018}
{Samui} S.,  {Subramanian} K.,   {Srianand } R.,  2018, \mn@doi [\mnras]
  {10.1093/mnras/sty287}, \href
  {https://ui.adsabs.harvard.edu/abs/2018MNRAS.476.1680S} {476, 1680}

\bibitem[\protect\citeauthoryear{{Schekochihin} \& {Cowley}}{{Schekochihin} \&
  {Cowley}}{2006}]{SchekochihinC2006}
{Schekochihin} A.~A.,  {Cowley} S.~C.,  2006, \mn@doi [Physics of Plasmas]
  {10.1063/1.2179053}, \href
  {https://ui.adsabs.harvard.edu/abs/2006PhPl...13e6501S} {13, 056501}

\bibitem[\protect\citeauthoryear{{Schekochihin} \& {Kulsrud}}{{Schekochihin} \&
  {Kulsrud}}{2001}]{SchekochihinK2001}
{Schekochihin} A.~A.,  {Kulsrud} R.~M.,  2001, \mn@doi [Physics of Plasmas]
  {10.1063/1.1404383}, \href
  {https://ui.adsabs.harvard.edu/abs/2001PhPl....8.4937S} {8, 4937}

\bibitem[\protect\citeauthoryear{{Schekochihin}, {Cowley}, {Taylor}, {Maron}
  \& {McWilliams}}{{Schekochihin} et~al.}{2004}]{SchekochihinEA2004}
{Schekochihin} A.~A.,  {Cowley} S.~C.,  {Taylor} S.~F.,  {Maron} J.~L.,
  {McWilliams} J.~C.,  2004, \mn@doi [\apj] {10.1086/422547}, \href
  {http://adsabs.harvard.edu/abs/2004ApJ...612..276S} {612, 276}

\bibitem[\protect\citeauthoryear{{Schlickeiser} \& {Shukla}}{{Schlickeiser} \&
  {Shukla}}{2003}]{SchlickeiserS2003}
{Schlickeiser} R.,  {Shukla} P.~K.,  2003, \mn@doi [\apjl] {10.1086/381246},
  \href {https://ui.adsabs.harvard.edu/abs/2003ApJ...599L..57S} {599, L57}

\bibitem[\protect\citeauthoryear{{Schober}, {Schleicher}, {Federrath},
  {Klessen}  \& {Banerjee}}{{Schober} et~al.}{2012a}]{SchoberEA2012a}
{Schober} J.,  {Schleicher} D.,  {Federrath} C.,  {Klessen} R.,   {Banerjee}
  R.,  2012a, \mn@doi [\pre] {10.1103/PhysRevE.85.026303}, \href
  {https://ui.adsabs.harvard.edu/abs/2012PhRvE..85b6303S} {85, 026303}

\bibitem[\protect\citeauthoryear{{Schober}, {Schleicher}, {Bovino}  \&
  {Klessen}}{{Schober} et~al.}{2012b}]{SchoberEA2012b}
{Schober} J.,  {Schleicher} D.,  {Bovino} S.,   {Klessen} R.~S.,  2012b,
  \mn@doi [\pre] {10.1103/PhysRevE.86.066412}, \href
  {https://ui.adsabs.harvard.edu/abs/2012PhRvE..86f6412S} {86, 066412}

\bibitem[\protect\citeauthoryear{{Schober}, {Schleicher}, {Federrath}, {Bovino}
   \& {Klessen}}{{Schober} et~al.}{2015}]{SchoberEA2015}
{Schober} J.,  {Schleicher} D.~R.~G.,  {Federrath} C.,  {Bovino} S.,
  {Klessen} R.~S.,  2015, \mn@doi [\pre] {10.1103/PhysRevE.92.023010}, \href
  {https://ui.adsabs.harvard.edu/abs/2015PhRvE..92b3010S} {92, 023010}

\bibitem[\protect\citeauthoryear{{Seta}}{{Seta}}{2019}]{Seta2019}
{Seta} A.,  2019, PhD thesis, Newcastle University, Newcastle Upon Tyne, UK,
  \url {http://theses.ncl.ac.uk/jspui/handle/10443/4685}

\bibitem[\protect\citeauthoryear{{Seta}, {Bhat}  \& {Subramanian}}{{Seta}
  et~al.}{2015}]{SetaBS2015}
{Seta} A.,  {Bhat} P.,   {Subramanian} K.,  2015, \mn@doi [Journal of Plasma
  Physics] {10.1017/S0022377815000628}, \href
  {http://adsabs.harvard.edu/abs/2015JPlPh..81e3903S} {81, 395810503}

\bibitem[\protect\citeauthoryear{{Seta}, {Shukurov}, {Wood}, {Bushby}  \&
  {Snodin}}{{Seta} et~al.}{2018}]{SetaEA2018}
{Seta} A.,  {Shukurov} A.,  {Wood} T.~S.,  {Bushby} P.~J.,   {Snodin} A.~P.,
  2018, \mn@doi [\mnras] {10.1093/mnras/stx2606}, \href
  {http://adsabs.harvard.edu/abs/2018MNRAS.473.4544S} {473, 4544}

\bibitem[\protect\citeauthoryear{Seta, Bushby, Shukurov  \& Wood}{Seta
  et~al.}{2020}]{SetaEA2020}
Seta A.,  Bushby P.~J.,  Shukurov A.,   Wood T.~S.,  2020, \mn@doi [Phys. Rev.
  Fluids] {10.1103/PhysRevFluids.5.043702}, 5, 043702

\bibitem[\protect\citeauthoryear{{Sharma}, {Jagannathan}, {Seshadri}  \&
  {Subramanian}}{{Sharma} et~al.}{2017}]{SharmaEA2017}
{Sharma} R.,  {Jagannathan} S.,  {Seshadri} T.~R.,   {Subramanian} K.,  2017,
  \mn@doi [\prd] {10.1103/PhysRevD.96.083511}, \href
  {https://ui.adsabs.harvard.edu/abs/2017PhRvD..96h3511S} {96, 083511}

\bibitem[\protect\citeauthoryear{{Sharma}, {Subramanian}  \&
  {Seshadri}}{{Sharma} et~al.}{2018}]{SharmaSS2018}
{Sharma} R.,  {Subramanian} K.,   {Seshadri} T.~R.,  2018, \mn@doi [\prd]
  {10.1103/PhysRevD.97.083503}, \href
  {https://ui.adsabs.harvard.edu/abs/2018PhRvD..97h3503S} {97, 083503}

\bibitem[\protect\citeauthoryear{{Shukurov}}{{Shukurov}}{2004}]{Shukurov2004}
{Shukurov} A.,  2004, ArXiv Astrophysics e-prints, \href
  {http://adsabs.harvard.edu/abs/2004astro.ph.11739S} {}

\bibitem[\protect\citeauthoryear{Shukurov \& Sokoloff}{Shukurov \&
  Sokoloff}{2008}]{ShukurovS2008}
Shukurov A.,  Sokoloff D.,  2008, in Cardin P.,  Cugliandolo L.,  eds, Les
  Houches, Vol.~88, Dynamos.
Elsevier, pp 251 -- 299

\bibitem[\protect\citeauthoryear{{Subramanian}}{{Subramanian}}{1999}]{Subramanian1999}
{Subramanian} K.,  1999, \mn@doi [\prl] {10.1103/PhysRevLett.83.2957}, \href
  {http://adsabs.harvard.edu/abs/1999PhRvL..83.2957S} {83, 2957}

\bibitem[\protect\citeauthoryear{{Subramanian}}{{Subramanian}}{2003}]{Subramanian2003}
{Subramanian} K.,  2003, \mn@doi [\prl] {10.1103/PhysRevLett.90.245003}, \href
  {https://ui.adsabs.harvard.edu/#abs/2003PhRvL..90x5003S} {90, 245003}

\bibitem[\protect\citeauthoryear{{Subramanian}}{{Subramanian}}{2016}]{Subramanian2016}
{Subramanian} K.,  2016, \mn@doi [Reports on Progress in Physics]
  {10.1088/0034-4885/79/7/076901}, \href
  {https://ui.adsabs.harvard.edu/abs/2016RPPh...79g6901S} {79, 076901}

\bibitem[\protect\citeauthoryear{{Subramanian}}{{Subramanian}}{2019}]{Subramanian2019}
{Subramanian} K.,  2019, \mn@doi [Galaxies] {10.3390/galaxies7020047}, \href
  {https://ui.adsabs.harvard.edu/abs/2019Galax...7...47S} {7, 47}

\bibitem[\protect\citeauthoryear{{Subramanian}, {Narasimha}  \&
  {Chitre}}{{Subramanian} et~al.}{1994}]{SubramanianNC1994}
{Subramanian} K.,  {Narasimha} D.,   {Chitre} S.~M.,  1994, \mn@doi [\mnras]
  {10.1093/mnras/271.1.L15}, \href
  {https://ui.adsabs.harvard.edu/abs/1994MNRAS.271L..15S} {271, L15}

\bibitem[\protect\citeauthoryear{{Sur}, {Federrath}, {Schleicher}, {Banerjee}
  \& {Klessen}}{{Sur} et~al.}{2012}]{SurEA2012}
{Sur} S.,  {Federrath} C.,  {Schleicher} D. R.~G.,  {Banerjee} R.,   {Klessen}
  R.~S.,  2012, \mn@doi [\mnras] {10.1111/j.1365-2966.2012.21100.x}, \href
  {https://ui.adsabs.harvard.edu/#abs/2012MNRAS.423.3148S} {423, 3148}

\bibitem[\protect\citeauthoryear{{Sur}, {Bhat}  \& {Subramanian}}{{Sur}
  et~al.}{2018}]{SurBS2018}
{Sur} S.,  {Bhat} P.,   {Subramanian} K.,  2018, \mn@doi [\mnras]
  {10.1093/mnrasl/sly007}, \href
  {https://ui.adsabs.harvard.edu/#abs/2018MNRAS.475L..72S} {475, L72}

\bibitem[\protect\citeauthoryear{{Tavecchio}, {Ghisellini}, {Bonnoli}  \&
  {Foschini}}{{Tavecchio} et~al.}{2011}]{TavecchioEA2011}
{Tavecchio} F.,  {Ghisellini} G.,  {Bonnoli} G.,   {Foschini} L.,  2011,
  \mn@doi [\mnras] {10.1111/j.1365-2966.2011.18657.x}, \href
  {https://ui.adsabs.harvard.edu/abs/2011MNRAS.414.3566T} {414, 3566}

\bibitem[\protect\citeauthoryear{{Tiede}, {Broderick}, {Shalaby}, {Pfrommer},
  {Puchwein}, {Chang}  \& {Lamberts}}{{Tiede} et~al.}{2020}]{TiedeEA2020}
{Tiede} P.,  {Broderick} A.~E.,  {Shalaby} M.,  {Pfrommer} C.,  {Puchwein} E.,
  {Chang} P.,   {Lamberts} A.,  2020, \mn@doi [\apj]
  {10.3847/1538-4357/ab737e}, \href
  {https://ui.adsabs.harvard.edu/abs/2020ApJ...892..123T} {892, 123}

\bibitem[\protect\citeauthoryear{{Tzeferacos} et~al.,}{{Tzeferacos}
  et~al.}{2018}]{TzeferacosEA2018}
{Tzeferacos} P.,  et~al., 2018, \mn@doi [Nature Communications]
  {10.1038/s41467-018-02953-2}, \href
  {https://ui.adsabs.harvard.edu/abs/2018NatCo...9..591T} {9, 591}

\bibitem[\protect\citeauthoryear{{Va{\u{i}}nshte{\u{i}}n} \&
  {Zel'dovich}}{{Va{\u{i}}nshte{\u{i}}n} \&
  {Zel'dovich}}{1972}]{VainshteinZ1972}
{Va{\u{i}}nshte{\u{i}}n} S.~I.,  {Zel'dovich} Y.~B.,  1972, \mn@doi [Soviet
  Physics Uspekhi] {10.1070/PU1972v015n02ABEH004960}, \href
  {https://ui.adsabs.harvard.edu/abs/1972SvPhU..15..159V} {15, 159}

\bibitem[\protect\citeauthoryear{{Vazza}, {Brunetti}, {Br{\"u}ggen}  \&
  {Bonafede}}{{Vazza} et~al.}{2018}]{VazzaEA2018}
{Vazza} F.,  {Brunetti} G.,  {Br{\"u}ggen} M.,   {Bonafede} A.,  2018, \mn@doi
  [\mnras] {10.1093/mnras/stx2830}, \href
  {https://ui.adsabs.harvard.edu/#abs/2018MNRAS.474.1672V} {474, 1672}

\bibitem[\protect\citeauthoryear{Waagan, Federrath  \& Klingenberg}{Waagan
  et~al.}{2011}]{WaaganFK2011}
Waagan K.,  Federrath C.,   Klingenberg C.,  2011, J. Comput. Phys., 230, 3331

\bibitem[\protect\citeauthoryear{{Weiss}}{{Weiss}}{1966}]{Weiss1966}
{Weiss} N.~O.,  1966, \mn@doi [Proceedings of the Royal Society of London
  Series A] {10.1098/rspa.1966.0173}, \href
  {https://ui.adsabs.harvard.edu/#abs/1966RSPSA.293..310W} {293, 310}

\bibitem[\protect\citeauthoryear{{Widrow}}{{Widrow}}{2002}]{Widrow2002}
{Widrow} L.~M.,  2002, \mn@doi [Reviews of Modern Physics]
  {10.1103/RevModPhys.74.775}, \href
  {http://adsabs.harvard.edu/abs/2002RvMP...74..775W} {74, 775}

\bibitem[\protect\citeauthoryear{{Widrow}, {Ryu}, {Schleicher}, {Subramanian},
  {Tsagas}  \& {Treumann}}{{Widrow} et~al.}{2012}]{WidrowEA2012}
{Widrow} L.~M.,  {Ryu} D.,  {Schleicher} D. R.~G.,  {Subramanian} K.,  {Tsagas}
  C.~G.,   {Treumann} R.~A.,  2012, \mn@doi [\ssr] {10.1007/s11214-011-9833-5},
  \href {https://ui.adsabs.harvard.edu/abs/2012SSRv..166...37W} {166, 37}

\bibitem[\protect\citeauthoryear{{Zeldovich}, {Ruzmaikin}  \&
  {Sokoloff}}{{Zeldovich} et~al.}{1990}]{ZeldovichRS1990}
{Zeldovich} {\relax Ya}.~B.,  {Ruzmaikin} A.~A.,   {Sokoloff} D.~D.,  1990,
  {The Almighty Chance}.
World Scientific, Singapore

\makeatother
\end{thebibliography}



\appendix

\bsp	
\label{lastpage}
\end{document}